\documentclass[draftcls,onecolumn,11pt,twoside]{IEEEtran}
\usepackage{cite}      % Written by Donald Arseneau
\usepackage{graphicx}  % Written by David Carlisle and Sebastian Rahtz
\usepackage{psfrag}    % Written by Craig Barratt, Michael C. Grant,
\usepackage{subfigure} % Written by Steven Douglas Cochran
\usepackage{url}       % Written by Donald Arseneau
\usepackage{amssymb}
\usepackage{amsmath}   % From the American Mathematical Society
\usepackage{array}
\newtheorem{theorem}{Theorem}

\newtheorem{lemma}{Lemma}

\newtheorem{remark}{Remark}

\newcommand{\sinc}{\ensuremath{\text{sinc}}}
\usepackage{xspace}
\newcommand{\bs}{\boldsymbol}
\newcommand{\Ddef}{\stackrel{\Delta}{=}}

\newcommand{\el}{\hfill\ensuremath{\triangle}}

\begin{document}

\title{Multiple-Description Coding by Dithered Delta-Sigma Quantization}

\author{Jan \O stergaard and Ram Zamir%
\thanks{
The material in this paper was presented in part at the IEEE Data Compression Conference, Snowbird, Utah, March 2007. The work of J.\ {\O}stergaard is supported by the Danish Research Council for Technology and Production Sciences, grant no.\ 274-07-0383.
J.\ \O stergaard (janoe@ieee.org) is with the Department of Electronic Systems, Aalborg University, Aalborg, Denmark.
R.\ Zamir (zamir@eng.tau.ac.il) is with the Department of Electrical Engineering-Systems, Tel Aviv University, Tel Aviv, Israel.}}

\maketitle

%
% note the % following the last \IEEEmembership and also the first \thanks -
% these prevent an unwanted space from occurring between the last author name
% and the end of the author line. i.e., if you had this:
%
% \author{....lastname \thanks{...} \thanks{...} }
%                     ^------------^------------^----Do not want these spaces!
%
% a space would be appended to the last name and could cause every name on that
% line to be shifted left slightly. This is one of those "LaTeX things". For
% instance, "A\textbf{} \textbf{}B" will typeset as "A B" not "AB". If you want
% "AB" then you have to do: "A\textbf{}\textbf{}B"
% \thanks is no different in this regard, so shield the last } of each \thanks
% that ends a line with a % and do not let a space in before the next \thanks.
% Spaces after \IEEEmembership other than the last one are OK (and needed) as
% you are supposed to have spaces between the names. For what it is worth,
% this is a minor point as most people would not even notice if the said evil
% space somehow managed to creep in.
%
% The paper headers
\markboth{IEEE Transactions on Information Theory, Vol. XX, No. XX, XXXX 2009}{\O stergaard and Zamir: Multiple-Description Coding by Dithered Delta-Sigma Quantization}
%
%\markboth{Journal of \LaTeX\ Class Files,~Vol.~1, No.~11,~November~2002}{Shell \MakeLowercase{\textit{et al.}}: Bare Demo of IEEEtran.cls for Journals}
% The only time the second header will appear is for the odd numbered pages
% after the title page when using the twoside option.
%
% *** Note that you probably will NOT want to include the author's name in ***
% *** the headers of peer review papers.                                   ***

% If you want to put a publisher's ID mark on the page
% (can leave text blank if you just want to see how the
% text height on the first page will be reduced by IEEE)
%\pubid{0000--0000/00\$00.00~\copyright~2002 IEEE}

% use only for invited papers
%\specialpapernotice{(Invited Paper)}

% make the title area
%\maketitle

\begin{abstract}
We address the connection between the multiple-description (MD) problem and Delta-Sigma quanti\-zation.
The inherent redundan\-cy due to oversampling in Delta-Sigma quantization, and the simple linear-additive noise model resulting from dithered lattice quantization, allow us to construct a symmetric and time-invariant MD coding scheme.
We show that the use of a noise shaping filter makes it possible to trade off central distortion for side distortion.
Asymp\-totically as the dimension of the lattice vector quantizer and order of the noise shaping filter approach infinity,
the entropy rate of the dithered Delta-Sigma quantization scheme approaches the symmetric two-channel MD rate-distortion function for a memoryless Gaussian source and MSE fidelity criterion, at any side-to-central distortion ratio and any resolution.
In the optimal scheme, the infinite-order noise shaping filter must be minimum phase and have a piece-wise flat power spectrum with a single jump discontinuity.
An important advantage of the proposed design is that it is symmetric in rate and distortion by construction, so the coding rates of the descriptions are identical and there is therefore no need for source splitting. 
\end{abstract}

% Note that keywords are not normally used for peerreview papers.

% For peer review papers, you can put extra information on the cover
% page as needed:
% \begin{center} \bfseries EDICS Category: 3-BBND \end{center}
%
% For peerreview papers, inserts a page break and creates the second title.
% Will be ignored for other modes.
%\IEEEpeerreviewmaketitle

\section{Introduction and Motivation}
% The very first letter is a 2 line initial drop letter followed
% by the rest of the first word in caps.
%
% form to use if the first word consists of a single letter:
% \PARstart{A}{demo} file is ....
%
% form to use if you need the single drop letter followed by
% normal text (unknown if ever used by IEEE):
% \PARstart{A}{}demo file is ....
%
% Some journals put the first two words in caps:
% \PARstart{T}{his demo} file is ....
%
\label{sec:intro}
\IEEEPARstart{D}{elta-Sigma} analogue to digital (A/D) conversion is a technique where the input signal is highly oversampled before being quantized by a low resolution quantizer. The quantization noise is then processed by a noise shaping filter which reduces the energy of the so-called in-band noise spectrum, i.e.\ the part of the noise spectrum which overlaps the spectrum of the input signal.
The end result is high bit-accuracy (A/D) conversion even in the presence of imperfections in the analogue components of the system, cf.~\cite{candy:1992}.

The process of oversampling and use of feedback to reduce quantization noise is not limited to A/D conversion of continuous-time signals but is in fact equally applicable to, for example, discrete time signals in which case we will use the term Delta-Sigma quantization.
Hence, given a discrete time signal we can apply Delta-Sigma quantization in order to discretize the amplitude of the signal and thereby obtain a digital signal.
It should be clear that the process of oversampling is not required in order to obtain a digital signal.
However, oversampling leads to a controlled amount of redundancy in the digital signal. This redundancy can be exploited in order to achieve a certain degree of robustness against inaccuracy in the quantization, 
or partial loss of information due to transmission of the digital signal over error-prone channels.
In this paper we pursue the latter aspect, and relate it to the problem of multiple descriptions.

In the information theory community the problem of quantization is usually referred to as a source coding problem whereas the problem of reliable transmission is referred to as a channel coding problem. Their combination then forms a joint source-channel coding problem. The multiple-description (MD) problem~\cite{elgamal:1982}, which has recently received a lot of attention, is basically a joint source-channel coding problem.
The MD problem is concerned with lossy encoding of information for transmission over an unreliable $K$-channel communication system. The channels may break down resulting in erasures and a loss of information at the receiving side.
The receiver knows which subset of the $K$ channels is working; the transmitter does not.
The problem is then to design an MD system which, for given channel rates, minimizes the distortions due to reconstruction of the source using information from any subsets of the channels.
Currently, the achievable MD rate-distortion region is only completely known for the case of two channels, squared-error fidelity criterion and a memoryless Gaussian source~\cite{ozarow:1980,elgamal:1982}. The bounds of~\cite{ozarow:1980} have been extended to stationary and smooth sources in~\cite{zamir:1999,zamir:2000}, where they were proven to be asymptotically tight at high resolution.
Inner and outer bounds to the rate-distortion region for the case of $K>2$ channels were presented in~\cite{venkataramani:2003,pradhan:2004,puri:2005} but it is not known whether any of the bounds are tight for $K>2$ channels.

The earliest practical MD schemes, which were shown to be asymptotically optimal at high resolution and large lattice vector quantizer dimensions, were based on the principle of index assignments, cf.~\cite{vaishampayan:1993,vaishampayan:2001,diggavi:2002,ostergaard:2004b,ostergaard:2007a}. Unfortunately, the existing methods for constructing index assignments in high vector dimensions are complex and computationally demanding.
To avoid the difficulty of designing efficient index assignments,
it was sug\-gested in~\cite{dayan:2002} that the index assignments of a two-description system can be replaced by successive quantization and linear estimation.
More specifically, the two side descriptions can be linearly combined and further enhanced by a refinement layer to yield the central reconstruction.
The design of~\cite{dayan:2002} suffers from a rate loss of 0.5 bit/dim.\ at high resolution and is therefore 
not able to achieve the MD rate-distortion bound.\footnote{The term \emph{rate loss} refers to the rate redundancy of the specific implementation, i.e.\ the additional rate required due to using a sub-optimal MD scheme.}
Recently, however, this gap was closed by Chen et al.~\cite{chen:2006} who
recognized that the rate region of the MD problem forms a polymatroid, and showed that the corner points of this rate region can be achieved by successive estimation and quantization.
The design of Chen et al.\ is inherently \emph{asymmetric} in the description rate since any corner point of a 
non-trivial rate region will lead to asymmetric rates.
To symmetrize the coding rates, it is necessary either to time-share between corner points, or
to break the quantization process into additional stages, which is a method known as ``source splitting'' (following Urbanke and Rimoldi's rate splitting approach for the multiple access channel).
When finite-dimensional quantizers are employed,
there is a space-filling loss due to the fact that the quantizer's Voronoi cells are finite dimensional and not
completely spherical,~\cite{gersho:1979},
and as such each description suffers a rate loss.
The rate loss of the design given in~\cite{chen:2006} is that of $2K-1$ quantizers because source splitting is performed by using an additional $K-1$ quantizers besides the conventional $K$ side quantizers.\footnote{%
By use of time-sharing, the rate loss can be reduced to that of only $K$ quantizers. Moreover, in the two-description scalar deterministic case, the rate loss can be further reduced, cf.~\cite{chen:2006}.}

An interesting open question is: can we avoid both the complexity of the index assignments and the extra space-filling loss due to source splitting in symmetric\footnote{By \emph{symmetric} we refer to the case where the MD scheme has balanced description rates and balanced side distortions.} MD coding?

Inspired by the works presented in~\cite{jayant:1981,dayan:2002,chen:2006}, we present a two-channel MD scheme based on two times oversampled dithered Delta-Sigma quantization, which is inherently symmetric in the description rate and as such there is no need for source splitting.\footnote{It should be noted that it is difficult to extend the proposed construction to allow for asymmetric description rates.}
The rate loss when employing finite-dimensional quantizers (in parallel) is therefore given by that of two quantizers.
The side-to-central distortion ratio is controlled by the noise shaping filter; the more ``high-pass'' the noise is, the larger is the side-to-central distortion ratio.
Asymp\-totically as the dimension of the vector quantizer and order of the noise shaping filter approach infinity, we show that the symmetric two-channel MD rate-distortion function for a memoryless Gaussian source and MSE fidelity criterion can be achieved at any resolution. 
It is worth emphasi\-zing that our design is not limited to two descriptions but, in fact, an arbitrary number of descriptions can be created simply by increasing the oversampling ratio.\footnote{%
When considering more than two descriptions, the distortion generally depends upon the particular subset of received descriptions whereas the coding rate is the same for all descriptions.
}
However, in this paper, we only prove optimality for the case of two descriptions.

In the Delta-Sigma quantization literature there seems to be a consensus of avoiding long feedback filters. We suspect this is mainly due to the fact that the quantization error in traditional Delta-Sigma quantization is a deterministic non-linear function of the input signal, which makes it difficult to perform an exact system analysis. Thus, there might be concerns regarding the stability of the system.
In our work we use dithered (lattice) quantization,
so that the quantization error is a stochastic process, independent of the input signal,
and the whole system becomes linear.
This linearization is highly desirable, since it allows an exact system analysis for any filter order and at any resolution.\footnote{Notice that our results are valid in steady state where the system is time invariant, i.e.\ we assume the system has been operating for a long time so that possible short-time temporal transient effects can be ignored. When referring to variances and power spectra we therefore always mean the \emph{stationary} variances and \emph{stationary} power spectra~\cite{costa:2005}.}
%For finite filter order, we show that the optimal filter coefficients are found by solving a set of Yule-Walker equations.
The case of infinite filter order has a very simple solution in the frequency domain,
which (for large lattice dimension) guarantees that the proposed scheme achieves the
symmetric two-channel MD rate-distortion function~\cite{ozarow:1980,elgamal:1982}.

% To gain some insight into why this solution is asymptotically optimal,
% observe that the Delta-Sigma quantization structure resembles the nature of the optimum test channel that achieves the two-channel MD rate-distortion region~\cite{ozarow:1980,elgamal:1982}.
% This channel (as shown in Fig.~\ref{fig:testchannel}) has two additive noise branches $U_0=X+N_0$ and $U_1=X+N_1$, where the pair $(N_0, N_1)$ is \emph{negatively} correlated.
% At high resolution conditions and symmetric rates and distortions, %this system is particularly simple;
% the side reconstructions $\hat{X}_0$ and $\hat{X}_1$ become $\hat{X}_0=U_0$ and $\hat{X}_1=U_1$, while the central reconstruction $\hat{X}_c$ becomes a simple average, i.e.\ $\hat{X}_c=(\hat{X}_0+\hat{X}_1)/2$.
% We may view the negatively correlated additive noises as adjacent
% samples of "highpass noise", and the averaging operation of
% the central reconstruction as "lowpass filtering".
% Intuitively, for a fixed side distortion the central distortion is reduced by
% shaping the spectrum of the noise to be away from the source
% band (the source component in $U_0$ and $U_1$ is the same
% which amounts to a lowpass signal).
% Thus, Delta-Sigma quantization provides a
% time-invariant filter version of this double branch test channel.
% This is further addressed in Section~\ref{sec:dsq_ozarow}.

Besides the quantizer-based MD schemes mentioned above there exist
several other approaches, e.g.\ MD schemes based on quantized
overcomplete
expansions~\cite{chou:1999,goyal:2001b,dragotti:2001,kovacevic:2002}.
The works of~\cite{chou:1999,goyal:2001b} are
based on finite frame expansions and that
of~\cite{dragotti:2001,kovacevic:2002} are based on redundant
$M$-channel filter banks.

It is well known that there is a connection between quantized overcomplete expansions and Delta-Sigma quantization, cf.~\cite{bolcskei:2001,boufounos:2005,boufounos:2006}. Furthermore, as mentioned above, the connection between overcomplete expansions and the MD problem has also been established.
Yet, to the best of the authors knowledge, none of the schemes presented in~\cite{chou:1999,goyal:2001b,dragotti:2001,kovacevic:2002} are able to achieve the above mentioned MD rate-distortion bounds. Furthermore, the use of Delta-Sigma quantization explicitly for MD coding appears to be a new idea. 
In this paper, we show that traditional Delta-Sigma quantization can be recast in the context of MD coding and furthermore, that it provides an optimal solution to the MD problem in the symmetric case.

The paper is structured as follows: 
In Section~\ref{sec:deltasigma}, we provide an introduction to dithered Delta-Sigma quantization. In 
Section~\ref{sec:mdchr}, a connection between Delta-Sigma quantization and MD coding is established and we present
 the main theorem of this work. 
The proof of the theorem is deferred to Section~\ref{sec:proofs}. 
Section~\ref{sec:asympdesc} presents an asymptotic characterization and performance analysis of the
proposed scheme in the limit of high dimensional vector quantization
and high order noise shaping filter. 
%The non-asymptotic analysis is given in Section~\ref{sec:nonasymptdesc}.
Section~\ref{sec:universal} shows that the proposed scheme is, in fact,
asymptotically optimal at high resolution for any i.i.d.\ source with finite differential entropy. 
An extension to $K$ descriptions is presented in Section~\ref{sec:extensionK}, and 
finally, Section~\ref{sec:conclusion} contains the conclusions.

\section{Dithered Delta-Sigma Quantization}
\label{sec:deltasigma}
Throughout this paper we will use upper case letters for stochastic variables and lower case letters
for their realizations. Infinite sequences and $L$-dimensional vectors will be typed in bold face.
We let $X \sim N(0,\sigma_X^2)$ denote a zero-mean Gaussian variable of variance $\sigma_X^2$,
and $\bs{X}=\{X_1,X_2,\dotsc\}$ denote an infinite sequence of independent copies of $X$.
Thus $\bs{X}$ is an i.i.d.\ (white) Gaussian process.
Moreover, $\bs{x}=\{x_1,x_2,\dotsc,\}$ denotes a realization of $\bs{X}$ where $x_k$ is the $k$th symbol of $\bs{x}$.

%%%%%%%%%%%%%%%%%%%%%%%%%%%%%%%%%%%%%%%%%%%%%%%%%%%%%%%%%%%%%%%%%%%%%%%%%%%%%%%%%%%%%%%%%%%
%%%     ECDQ FIRST

\subsection{Preliminaries: Entropy-Coded Dithered Quantization}\label{sec:ecdq}
\begin{figure}[th]
\psfrag{X}{$\bs{S}$} \psfrag{Z}{$\bs{Z}$}
\psfrag{Xh}{$\bs{\hat{S}}$} \psfrag{Q}{\hspace{-2mm}$Q_L$}
\begin{center}
\includegraphics[width=8.9cm]{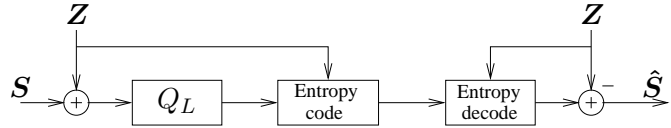}
\caption{Entropy-constrained dithered (lattice) quantization
(ECDQ). The dither signal $\bs{Z}$ is assumed known at the
decoder. The quantizer $Q_L$ is an $L$-dimensional lattice vector
quantizer and the rate of the entropy coder is given by the
entropy of the quantized output of $Q_L$ conditioned upon
$\bs{Z}$.} \label{fig:ecdq}
\end{center}
\vspace{0mm}
\end{figure}

Before introducing our dithered Delta-Sigma quantization system,
let us recall the properties of entropy-coded dithered
(lattice) quan\-tization (ECDQ)~\cite{zamir:1992}.
ECDQ relies upon subtractive dither; see Fig.~\ref{fig:ecdq}.
For an $L$-dimensional input vector $\bs{S}$, the ECDQ output is given by
$\bs{\hat{S}} = Q_L(\bs{S} + \bs{Z}) - \bs{Z}$, where $Q_L$
denotes an $L$-dimensional lattice quantizer with Voronoi cells
\cite{conway:1999}.
The dither vector $\bs{Z}$, which is known to both the encoder and
the decoder, is independent of the input signal and previous
realizations of the dither, and is uniformly distributed over the
basic Voronoi cell of the lattice quantizer.
It follows that the quantization error
\begin{equation}
\label{eq:quantizationerror}
\bs{E} = \bs{\hat{S}}-\bs{S} = Q_L(\bs{S} + \bs{Z}) - \bs{S}
-\bs{Z}
\end{equation}
is statistically independent of the input signal. Furthermore,
$\bs{E}$ is an i.i.d.-vector process, where each $L$-block is
uniformly distributed over the mirror image of the basic cell of
the lattice, i.e., as $-\bs{Z}$.
In particular, it follows that $\bs{E}$
is a zero-mean white vector with variance $\sigma_{E}^2$~\cite{zamir:1992,zamir:1996}.

The average code length of the quantized variables is given by the
conditional entropy $H(Q_L(\bs{S}+\bs{Z})|\bs{Z})$ of the
quantizer $Q_L$, where the conditioning is with respect to the
dither vector $\bs{Z}$. It is known that this conditional
entropy is equal to the mutual information over the additive
noise channel $\bs{Y} = \bs{S}+\bs{E}$ where $\bs{E}$ (the
channel's noise) is distributed as $-\bs{Z}$;
see~\cite{zamir:1992} for details. The coding rate (per $L$-block)
of the quantizer is therefore given by
\begin{eqnarray}\label{eq:ECDQrate}
H(Q_L(\bs{S}+\bs{Z})|\bs{Z}) =
I(\bs{S};\bs{Y}) = h(\bs{S}+\bs{E}) - h(\bs{E})
\end{eqnarray}
where $I(\cdot,\cdot)$ denotes the mutual information and $h(\cdot)$ denotes 
the differential entropy.
If subsequent quantizer outputs are entropy-coded jointly, then we must
change the blockwise mutual information in the rate formula (\ref{eq:ECDQrate}) to
the joint mutual information between input-output sequences (if there is no feedback)
\cite{zamir:1992},
or to the \emph{directed} mutual information (if there is feedback)
\cite{massey:1990,zamir:2007}. 

If the source $\bs{S}$ is white Gaussian, then the coding rate
(\ref{eq:ECDQrate}), normalized per-sample, is upper bounded by
%\begin{eqnarray}
\begin{align} \notag
\frac{1}{L}H(Q_L(\bs{S}+\bs{Z})|\bs{Z}) &\leq
\frac{1}{2}\log_2\left( 1 +
\frac{\text{Var}(S_k)}{\sigma_E^2}\right)  \\ \notag
&\quad+ \frac{1}{2}\log_2(2\pi e G_L) \\
\label{eq:rateecdq} &= R_S(D) + \frac{1}{2}\log_2(2\pi e G_L)
\end{align}
where $G_L$ is the dimensionless normalized second moment of the
$L$-dimensional lattice quantizer $Q_L$~\cite{conway:1999}. In the
second equality, $D$ is the total distortion after a suitable
post-filter (multiplier) and $R_S(D)$ is the rate-distortion
function of the white Gaussian source $S$; see \cite{zamir:1996b}.
The quantity $2\pi eG_L$ is the space-filling loss of the
quantizer and $\frac{1}{2}\log_2(2\pi e G_L)$ is the divergence of
the quantization noise from Gaussianity. It follows that it is
desirable to have Gaussian distributed quantization noise in order
to make $G_L$ as small as possible and thereby drive the rate of
the filtered quantizer towards $R_S(D)$. Fortunately, it is known
that there exists lattices where $G_L \rightarrow 1/2\pi e$ as
$L\rightarrow \infty$; the quantization noise of such
quantizers is white, and becomes asymptotically (in dimension)
Gaussian distributed in the divergence sense~\cite{zamir:1996}.

%\subsection{Comparison with a Single-Description System with Joint Entropy Coding}\label{sec:descriptionrate}

%We previously saw that, in the SD case, only the in-band noise spectrum affects the distortion,
%whereas the complete noise spectrum affects the MD distortions.
%Specifically, the in-band noise spectrum determines the central distortion and the sum of the
%in-band and out-of-band noise spectrum determines the side distortion.
%We can show a similar relationship with respect to the rates.
%First, notice from~(\ref{eq:descrate}) that the description rate in the MD case
%depends upon the complete noise spectrum.\footnote
%{
%Recall that the side distortion is associated with the entire noise spectrum through the
%sum $(\delta +\delta^{-1})/2$. This sum is also part of the rate expression in~(\ref{eq:descrate}).
%}

%%%%%%%%%%%%%%%%%%%%%%%%%%

%%%%%%%%%%%%%%%%%%%%%%%%%%%%%% END ECDQ %%%%%%%%%%%%%%%%%%%%%%%%%%%%%%%%%%%%%

%%%%%%%%%%%%%%%%%%%%%  DSQ scheme description %%%%%%%%%%%%%%%%%%%%%%%%%%%%%

\subsection{Delta-Sigma ECDQ}\label{sec:DSQ}
%
%\vspace{-5mm}
\begin{figure*}[th]
\psfrag{ak}{$a_k$}
\psfrag{ak'}{$a'_k$}
\psfrag{akh}{$\hat{a}_k$}
\psfrag{c(z)}{$c'(z)$}
\psfrag{Q}{$Q_L$}
\psfrag{ek}{$e_k$}
\psfrag{ekt}{$\tilde{e}_k$}
\psfrag{xn}{$x_n$}
\psfrag{xn1}{$x_{n'}$}
\psfrag{xh}{$\hat{x}_n$}
\psfrag{zk}{\small Dither}
\psfrag{h(z)}{$h(z)$}
\psfrag{ha(z)}{$h_a(z)$}
\psfrag{2}{$2$}
\psfrag{R(QLZ)}{\small $R=H(Q_L|\text{Dither})$}
\begin{center}
\includegraphics[width=15cm]{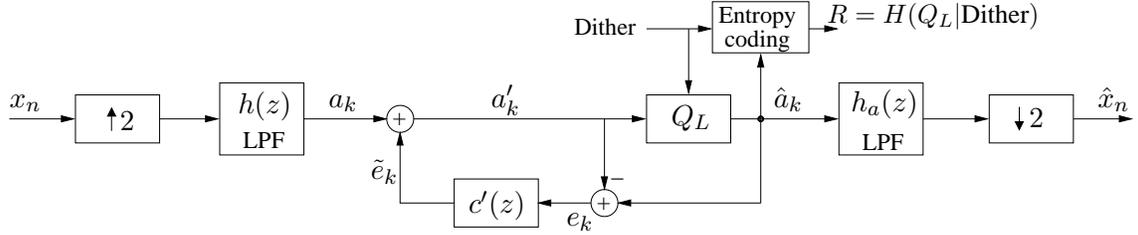}
\caption{Dithered Delta-Sigma quantization.}
\label{fig:deltasigma}
\end{center}
%\vspace{-8mm}
\end{figure*}

\begin{figure}[th]
\psfrag{SX}{\small $S_X$}
\psfrag{pi}{\small $\pi$}
\psfrag{-pi}{\small $-\pi$}
\psfrag{SA}{\small $S_A$}
\psfrag{pi2}{\small $\pi/2$}
\psfrag{-pi2}{\small $-\pi/2$}
\psfrag{w}{\small $\omega$}
\psfrag{0}{\small $0$}
\psfrag{A}{\small $A$}
\psfrag{X}{\small $X$}
\psfrag{2}{\small $2$}
\psfrag{s}{$\scriptstyle \sigma_X^2$}
\psfrag{2s}{$\scriptstyle 2\sigma_X^2$}
\begin{center}
\mbox{%
\subfigure[Spectrum of $X$]{\includegraphics[width=4.8cm]{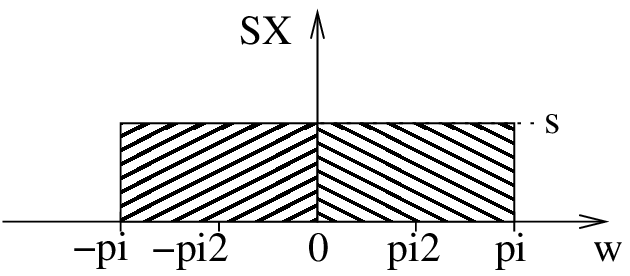}}
\subfigure[Oversampling by two]{\raisebox{5mm}{\includegraphics[width=3.5cm]{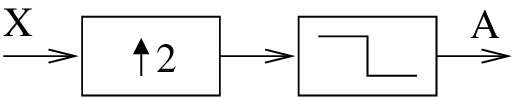}}}}\par
\subfigure[Spectrum of $A$]{\includegraphics[width=4.8cm]{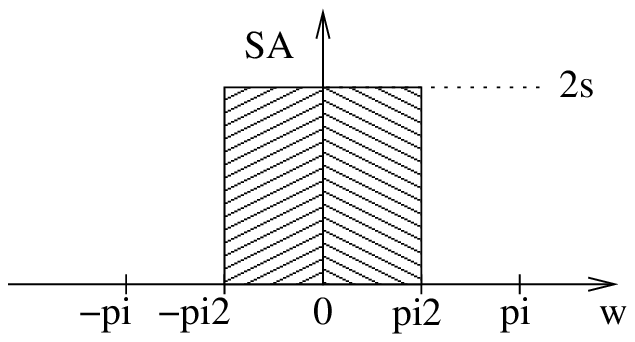}}
\caption{The power spectrum of (a) the input signal and (c) the oversampled signal. (b) illustrates the oversampling process where the input signal is first upsampled by two and then filtered by an ideal half-band lowpass filter.}
\label{fig:spectrum_X_A}
\end{center}
%\vspace{-5mm}
\end{figure}

We are now ready to introduce our dithered Delta-Sigma quantization system.
We begin with the single description case which is sketched in Fig.~\ref{fig:deltasigma}.\footnote{%
The Delta-Sigma quantization system shown in Fig.~\ref{fig:deltasigma} is a discrete-time version of
the \emph{general noise-shaping coder} presented in~\cite{tewksbury:1978}.
The system has an equivalent form where the feedback is first subtracted and this difference is then
filtered~\cite{tewksbury:1978}.}
The multiple descriptions case will be discussed in Section~\ref{sec:mdchr}.
The input sequence $\bs{x}$ is first oversampled by a factor of two to produce
the oversampled sequence $\bs{a}$.
It follows that $\bs{a}$ is a redundant representation of the input sequence $\bs{x}$,
which can be obtained simply by inserting a zero between every sample of $\bs{x}$ and applying an interpolating
(ideal lowpass) filter $h(z)$.
For a wide-sense stationary input process $\bs{X}$, the resulting oversampled signal $\bs{A}$
would be wide-sense stationary, with the same variance as the input, and the same
power-spectrum only squeezed to half the frequency band as shown in Fig.~\ref{fig:spectrum_X_A}.
In particular, a white Gaussian input becomes a half-band
low-pass Gaussian process with
\begin{equation}\label{eq:var_preserve}
 \mbox{Var}(A_k) = \mbox{Var}(X_k) = \sigma_X^2.
\end{equation}
At the other end of the system we apply an anti-aliasing filter $h_a(z)$,
i.e.\ an ideal half-band lowpass filter, and downsample by two in order to get back to the original sampling rate.

We would like to emphasize that the dithered Delta-Sigma quantization scheme
is not limited to oversampling ratios of two.
In fact, arbitrary (even fractional) oversampling ratios may be used.
This option is discussed further in Section~\ref{sec:extensionK}.

%%%%%%%%%%%%%%%%%%%%%%%%%%%%%%%%%%%%%%%%%%%%%%%%%%%%%%%%%%%%%%%%%%%%%%%
%%%   Noise Feedback (ECDQ was moved earlier!)
%%%%%%%%%%%%%%%%%%%%%%%%%%%%%%%%%%%%%%%%%%%%%%%%%%%%%%%%%%%%%%%%%%%%%%%%

The oversampled source sequence $\bs{a}$ is
combined with noise feedback $\bs{\tilde{e}}$,
and the resulting signal $\bs{a'}$ is sequentially quantized
on a sample by sample basis using a dithered quantizer.
For the simplicity of the exposition we shall momentarily assume
scalar quantization, i.e., $L=1$.  The extension to $L>1$ is discussed
in Section~\ref{sec:Vector_DSQ}.
The quantization error $e_k$
of the $k$th sample,
given for a general ECDQ by (\ref{eq:quantizationerror}),
is fed back through the (causal) filter $c'(z)=\sum_{i=1}^{p}c_iz^{-i}$
and combined with the next source sample $a_{k+1}$
to produce the next ECDQ input $a'_{k+1}$.
Thus, the output of the quantizer can be written as
\begin{equation}\label{eq:errors}
\hat{a}_k = a'_k + e_k = a_k + \tilde{e}_k + e_k \Ddef a_k + \epsilon_k
\end{equation}
where
$\tilde{e}(z)=c'(z)e(z)$ or equivalently
$$ \tilde{e}_k = \sum_{i=1}^{p}c_i e_{k-i}. $$

%%%%%%%%%%%%%%%%%%% Equivalent additive noise DSQ %%%%%%%%%%%

%
\begin{figure*}[th]
\psfrag{ak}{$a_k$}
\psfrag{ak'}{$a'_k$}
\psfrag{akh}{$\hat{a}_k$}
\psfrag{c(z)}{$c'(z)$}
\psfrag{Q}{$Q_L$}
\psfrag{ek}{$e_k$}
\psfrag{ekt}{$\tilde{e}_k$}
\psfrag{xn}{$x_n$}
\psfrag{xn1}{$\hat{x}_{n'}$}
\psfrag{xh}{$\hat{x}_n$}
\psfrag{ek}{$e_k$}
\psfrag{h(z)}{$h(z)$}
\psfrag{ha(z)}{$h_a(z)$}
\psfrag{r}{$2$}
\begin{center}
\includegraphics[width=15cm]{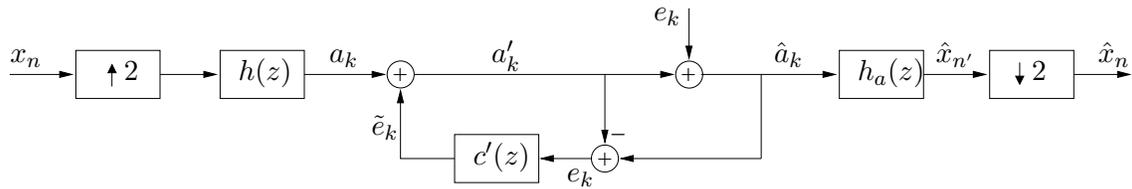}
\caption{The dithered quantizer is replaced by the additive noise model.}
\label{fig:deltasigma_model}
\end{center}
%\vspace{0mm}
\end{figure*}

\begin{figure}[th]
\psfrag{ak'}{$a'_k$}
\psfrag{Q}{$Q_L$}
\psfrag{ek}{$\!e_k$}
\psfrag{ekt}{$\tilde{e}_k$}
\psfrag{xn}{$x_n$}
\psfrag{xn1}{$\hat{x}_{n'}$}
\psfrag{xh}{$\hat{x}_n$}
\psfrag{h(z)}{$h(z)$}
\psfrag{ha(z)}{$\!\! h_a(z)$}
\psfrag{r}{$2$}
\psfrag{ak}{$a_k$}
\psfrag{akh}{$\hat{a}_k$}
\psfrag{c(z)}{$\!c(z)$}
\psfrag{es}{$\!\epsilon_k$}
\begin{center}
\includegraphics[width=9cm]{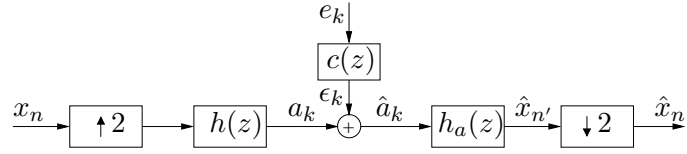}
\caption{The equivalent additive noise channel: The output $\hat{a}_k$ is obtained by passing the quantization error $e_k$ through the noise shaping filter $c(z)$ and adding the result to the input $a_k$.}
\label{fig:additivenoisechannel}
\end{center}
%\vspace{-5mm}
\end{figure}

As explained above,
the additive noise model is exact for ECDQ and we can therefore
represent the quantization operation as an additive noise channel, as
shown in Fig.~\ref{fig:deltasigma_model}.
In view of this linear model,
the equivalent reconstruction error in the oversampled domain,
denoted $\epsilon_k$ in (\ref{eq:errors}),
%% $$\epsilon_k=\hat{a}_k-a_k = e_k + \tilde{e}_k .$$
is statistically independent of the source.
Thus we call $\epsilon_k$ the ``equivalent noise''.
Notice that $\epsilon_k$ is obtained by passing the quantization
error $e_k$ through the equivalent
$p$th order noise shaping filter $c(z)$,
\begin{equation}\label{eq:noiseshapingfilter}
c(z)\triangleq  \sum_{i=0}^{p}c_iz^{-i}
\end{equation}
where $c_0=1$ so that $c(z)= 1 + c'(z)$.
%To see this, notice that the output is $\hat{a}(z) = a(z) + e(z) + c'(z)e(z)$,
%and the reconstruction error is therefore given by $\epsilon(z)=c(z)e(z)$,
%cf.\ Fig.~\ref{fig:equivnoise}.
Since the quantization error ${\bs e}$ of the ECDQ (\ref{eq:quantizationerror})
is white with variance $\sigma^2_E$,
it follows that the equivalent noise spectrum is given by
\begin{equation} \label{eq:equivalent_spectrum}
S_{\epsilon}(w) = |c(e^{jw})|^2 \sigma^2_E .
\end{equation}
The fact that the output $\hat{a}_k$ is obtained by passing the quantization
error $e_k$ through the noise shaping filter $c(z)$ and adding the result to
the input $a_k$ can be illustrated using an equivalent additive noise channel
as shown in Fig.~\ref{fig:additivenoisechannel}.

%\begin{figure}[th]
%\psfrag{eps}{$\epsilon_k$}
%\psfrag{c*(z)}{$c'(z)$}
%\psfrag{ek}{$e_k$}
%\begin{center}
%\includegraphics[width=5cm]{figs/equivnoise.eps}
%\caption{The reconstruction error $\epsilon_k=\tilde{e}_k + e_k$ is also called the ``equivalent noise'', since it can be obtained by passing the quantization error $e_k$ through the ``equivalent'' noise shaping filter $c(z)=1+c'(z)$.}
%\label{fig:equivnoise}
%\end{center}
%\end{figure}

%%%%%%%%%%%%%%%%%%%%  Noise Shaping %%%%%%%%%%%%%%%%%%%%%%%

\begin{figure}[th]
\psfrag{pi}{\small $\pi$}
\psfrag{-pi}{\small $-\pi$}
\psfrag{SA}{\small $S_A$}
\psfrag{SE}{\small $\sigma_E^2$}
\psfrag{pi2}{\hspace{1mm}\small \raisebox{-1mm}{$\frac{\pi}2$}}
\psfrag{-pi2}{\hspace{1mm}\small \raisebox{-1mm}{\hspace{1mm}$-\frac{\pi}2$}}
\psfrag{w}{\small $\omega$}
\psfrag{0}{\small $0$}
\psfrag{A}{\small $A$}
\psfrag{X}{\small $X$}
\psfrag{2}{\small $2$}
\psfrag{s}{$\scriptstyle 2\sigma_X^2$}
\psfrag{q}{$\scriptstyle \sigma_E^2$}
\psfrag{SCA}{$\!|c(e^{jw})|^2\sigma_E^2$}
\begin{center}
\hspace{1mm}
\mbox{\includegraphics[width=4.4cm]{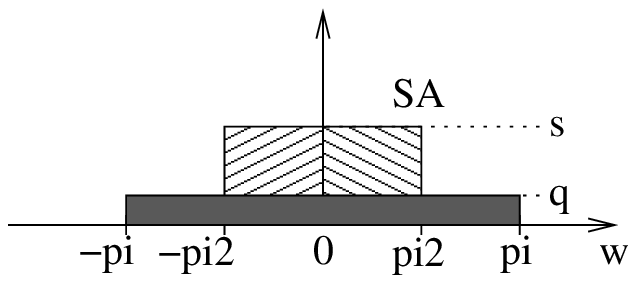}
\includegraphics[width=4.4cm]{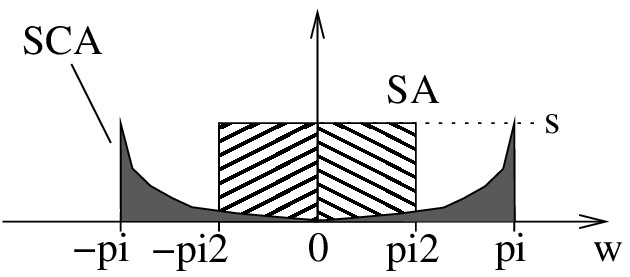}}
%\vspace{-5mm}
\caption{Illustrated on the left the case where there is no feedback and the quantization noise is therefore flat (in fact white) throughout the entire frequency range. On the right an example of noise shaping is illustrated. The grey-shaded areas illustrate the power spectra of the noise and the hatched areas illustrate the power spectra of the source.}
\label{fig:noiseshaping}
\end{center}
%\vspace{-5mm}
\end{figure}

We may view the feedback filter $c'(z)$ as if its
purpose is to predict the ``in-band'' noise component of $\tilde{e}_k$
based on the past $p$ quantization error samples
$e_{k-1}, e_{k-2}, \dotsc, e_{k-p}$ (at the expense of possibly increasing the ``out-of-band'' noise component). 
%%  as shown in Fig.~\ref{fig:deltasigma}.
%
The end result is that the equivalent noise spectrum (\ref{eq:equivalent_spectrum})
% \[
% S_{\epsilon}(w) = |c(e^{jw})|^2 \sigma^2_E
%  \]
is shaped away from the in-band part of the spectrum,
i.e., from the frequency range $(-\pi/2,+\pi/2)$,
as shown in Fig.~\ref{fig:noiseshaping}.
Notice that due to the anti-aliasing filter $h_a(z)$,
only the in-band noise
determines the overall system distortion.
The exact guidelines for this noise shaping are different in the single- and the
multiple description cases,
and will become clear in the sequel.
%
%\vspace{-3mm}

%%%%%%%%%%%%%%%%%%%%%%%%%%%%%%%%%%%%%%%

As previously mentioned,
if we encode the quantizer output symbols independently,
then the rate $R$ of the ECDQ is given by the mutual information between the input and the output of the
quantizer.
Thus, the rate (per sample) is given by
\begin{equation}\label{eq:ratesd1}
R = I(A_k' ; \hat{A}_k ) = I(A_k' ; A_k' + E_k)
\end{equation}
where $E_k$ is independent of the present and past
samples of $A_k'$ by the dithered quantization assumption.
If $A_k$ and $E_k$ were Gaussian, as discussed in Section~\ref{sec:Vector_DSQ} below,
then we could get
\begin{equation}\label{eq:ratesd2}
R = \frac{1}{2}\log_2\left( 1 + \frac{\text{Var}(A_k')}{\sigma_E^2}\right)
\end{equation}
where $\text{Var}(A_k')$ denotes the variance of the random variable $A_k'$.
At high resolution conditions the variance of the error
signal $\bs{e}$ (and therefore of $\bs{\epsilon}$)
is small compared to the source,
so by (\ref{eq:var_preserve})  we have
$\text{Var}(A'_k)  \approx \sigma_X^2$
which implies that (\ref{eq:ratesd2}) becomes
\begin{equation}\label{eq:ratesdhr}
R \approx \frac{1}{2}\log_2\left( \frac{\sigma_X^2}{\sigma_E^2}\right)
\end{equation}
where $\approx$ in (\ref{eq:ratesdhr}) is in the sense
that the difference between both sides of the equation goes to zero as
$\sigma_E^2 \rightarrow 0$.
We can now combine~(\ref{eq:ratesdhr}) with the expression~(\ref{eq:equivalent_spectrum}) for the noise spectrum to obtain a simple overall rate-distortion characterization of the system. It can be observed that the resulting $R(D)$ curve depends on both the in-band and the out-of-band noise components.

If we apply {\em joint} entropy coding of the quantizer outputs,
that is, we let the entropy coder take advantage of the memory
inside the oversampled source, then the rate of the Delta-Sigma
quantization scheme is independent of the out-of-band noise spectrum. To see this, 
recall that for jointly-coded ECDQ within a feedback loop,
the coding rate is given by the \emph{directed} mutual information rate, that is, \cite{zamir:2007},
\begin{align}\notag
&\bar{I}(A'_k \rightarrow A'_k + E_k) \\ \notag
&\overset{\hphantom{(a)}}{=} I(A'_k; A'_k + E_k| A'_{k-1} + E_{k-1}, A'_{k-2} + E_{k-2}, \dotsc ) \\ \notag
&\overset{\hphantom{(a)}}{=} h(A'_k + E_k | A'_{k-1} + E_{k-1},  A'_{k-2} + E_{k-2},\dotsc) - h(E_k) \\ \notag
&\overset{(a)}{=} h(A_k + \boldsymbol{\epsilon}_k | A_{k-1} + \boldsymbol{\epsilon}_{k-1}, A_{k-2} + \boldsymbol{\epsilon}_{k-2},\dotsc) - h(E_k) \\ \notag
&\overset{(b)}{=} \bar{h}(A + \boldsymbol{\epsilon}) - \bar{h}(\boldsymbol{\epsilon}) \\ \label{eq:directed}
&\overset{\hphantom{(a)}}{=} \bar{I}(A;A+\boldsymbol{\epsilon})
\end{align}
where $\bar{h}(\cdot)$ and $\bar{I}(\cdot)$ denote the entropy rate and mutual information rate, respectively. In the equations above $(a)$ follows since $A'_k=A_k + \tilde{E}_k$ and $\boldsymbol{\epsilon}_k=\tilde{E}_k+E_k$. In $(b)$ we used the fact that $E_k$ is the prediction error of $\boldsymbol{\epsilon}_k$ given its past so that $h(E_k)$ is the entropy rate of $\boldsymbol{\epsilon}$, i.e.\ $\bar{h}(\boldsymbol{\epsilon})=\bar{h}(E_k)=h(E_k)$.
Asymptotically as $L\rightarrow\infty$, the quantization noise becomes approximately Gaussian distributed, and the equivalent ECDQ channel in Fig.~\ref{fig:deltasigma_model} is AWGN (see Section~\ref{sec:Vector_DSQ} below). Recall that, for a Gaussian process, disjoint frequency bands are statistically independent. 
Therefore, since the input $A$ is lowpass, the mutual-information rate~(\ref{eq:directed}) is independent of the out-of-band part of the noise process $\boldsymbol{\epsilon}$. Thus, the joint-entropy coding rate is independent of the out-of-band noise spectrum.\footnote{Interestingly as we shall see later, in the MD case the coding rate depends on the in-band as well as the out-of-band noise spectra; see~(\ref{eq:descrate}).}

%%%%%%%%%%%%%%%%%%%%%%%%%%%%%%%%%%%%%%%%%%%%%%%%%%%%%%%%%%%%%%%%%%%%%%%%%%%%%%%%%%%%%%%%

\subsection{Vector Delta-Sigma Quantization}\label{sec:Vector_DSQ}
%
% It should be clear from the discussion about ECDQ
% in Section~\ref{sec:ecdq}
% that we would like to use high-dimensional quantizers, so that the
% quantization noise in (\ref{eq:ratesd1})
% is indeed approximately Gaussian.
% However, at first sight,
% it might appear as the sequential scalar nature of Delta-Sigma quantization prevents the use of anything but scalar quantizers. That this is not so will soon become clear. First, let us consider the scalar case, i.e.\ $L=1$.
% The input to the quantizer is $a'_k = a_k + \sum_{i=1}^p c_i e_{k-i}$
% and the output is $\hat{a}_k=a_k + \sum_{i=0}^p c_i e_{k-i}$.
% Since $a'_k$ is a scalar, the input to the quantizer is a scalar and the
% quantizer depicted in Fig.~\ref{fig:deltasigma} is therefore a scalar quantizer.

\begin{figure}[th]
\psfrag{ak}{$a^{(1)}_k$}
\psfrag{ak'}{$a'^{(1)}_k$}
\psfrag{akh}{\raisebox{0.5mm}{$\hat{a}_k^{(1)}$}}
\psfrag{c(z)}{$c'(z)$}
\psfrag{Q1}{\hspace{1mm}$Q_3^{(1)}$}
\psfrag{Q2}{\hspace{1mm}$Q_3^{(2)}$}
\psfrag{Q3}{\hspace{1mm}$Q_3^{(3)}$}
\psfrag{ek}{$\!e_k^{(1)}$}
\psfrag{ekt}{$\!\tilde{e}_k^{(1)}$}
\psfrag{ak2}{$a^{(2)}_k$}
\psfrag{ak'2}{$a'^{(2)}_k$}
\psfrag{akh2}{\raisebox{0.5mm}{$\hat{a}_k^{(2)}$}}
\psfrag{ek2}{$\!e_k^{(2)}$}
\psfrag{ekt2}{$\!\tilde{e}_k^{(2)}$}
\psfrag{ak3}{$a^{(3)}_k$}
\psfrag{ak'3}{$a'^{(3)}_k$}
\psfrag{akh3}{\raisebox{0.5mm}{$\hat{a}_k^{(3)}$}}
\psfrag{ek3}{$\!e_k^{(3)}$}
\psfrag{ekt3}{$\!\tilde{e}_k^{(3)}$}
\psfrag{xn}{$x_n$}
\psfrag{xn1}{$x_{n'}^{(1)}$}
\psfrag{xn2}{$x_{n'}^{(2)}$}
\psfrag{xn3}{$x_{n'}^{(3)}$}
\psfrag{ak1}{$a_k^{(1)}$}
\psfrag{Q3'}{$Q_3$}
%\psfrag{ak2}{$a_k^{(2)}$}
\psfrag{h(z)}{$h(z)$}
\begin{center}
\subfigure[Demultiplexing the i.i.d.\ source into $L=3$ independent streams]{\raisebox{0cm}{\includegraphics[width=5.5cm]{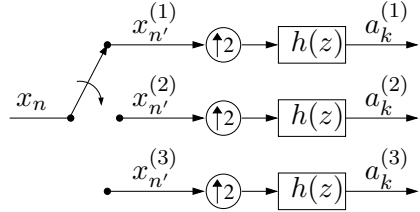}}}\par
\subfigure[Applying a three-dimensional lattice quantizer $Q_3$]{\includegraphics[width=7cm]{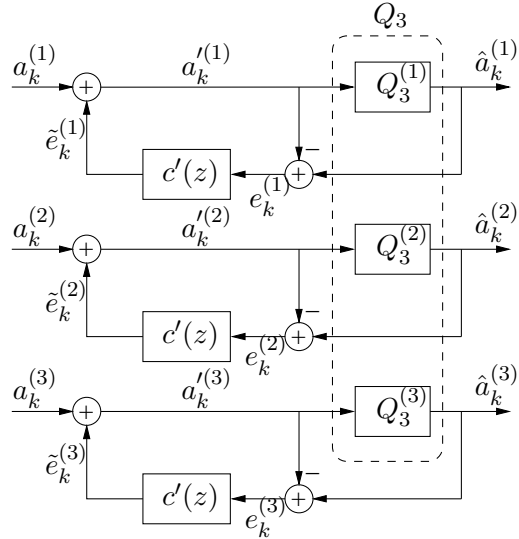}}
\caption{The dashed box illustrates that the triplet of scalars $(a'^{(1)}_k,a'^{(2)}_k,a'^{(3)}_k)$ are jointly quantized using three-dimensional ECDQ. Notice that we may see the three-dimensional lattice quantizer $Q_3$ as a composition of three functions where $\hat{a}^{(1)}_k=Q_3^{(1)}(a'^{(1)}_k,a'^{(2)}_k,a'^{(3)}_k),\ \hat{a}^{(2)}_k=Q_3^{(2)}(a'^{(1)}_k,a'^{(2)}_k,a'^{(3)}_k)$ and $\hat{a}^{(3)}_k=Q_3^{(3)}(a'^{(1)}_k,a'^{(2)}_k,a'^{(3)}_k)$.}
\label{fig:deltasigma_many}
\end{center}
%\vspace{-11mm}
\end{figure}

To justify the use of high-dimensional vector quantizers we will consider
a setup involving $L$ independent sources.\footnote{
The idea of applying lattice ECDQ to feedback coding systems in parallel was first
presented in~\cite{zamir:2007}.}
These sources can, for example, be obtained by demultiplexing the original memoryless process $X$ into $L$
independent parallel i.i.d.\ processes $X^{(l)}=\{X_{nL+l}\}, \forall n\in\mathbb{Z}$ and $l=1,\dotsc,L$.\footnote{Notice that the delay between two consecutive samples of the $l$th process will be that of $L$ input samples.} In this case the $n$th sample of the $l$th process $X^{(l)}$ is identical to the $(n\times L+l)$th sample of the original process $X$.
%In the case where $L=2$ we have two independent scalar processes, where $X^{(1)}$ consists of the even samples of $X$ and $X^{(2)}$ consists of the odd samples of $X$. 
Let us give an example where $L=3$ so that we have three processes $X^{(1)}, X^{(2)},$ and  $X^{(3)}$.
The three processes are each upsampled by a factor of two so that we obtain the three processes
$A^{(1)}, A^{(2)}$, and $A^{(3)}$, where each is input to a Delta-Sigma quantization system
as shown in Fig.~\ref{fig:deltasigma_many}.
Hence, in this case, three coders are operating in parallel and instead of a single sample $a'_k$ we have a triplet of independent samples $(a'^{(1)}_k,a'^{(2)}_k,a'^{(3)}_k)$.
This makes it possible to apply three-dimensional ECDQ on the vector formed by cascading the triplet of scalars.
If $L$ coders are operating in parallel, we can form the set of $L$ independent samples $(a'^{(1)}_k,a'^{(2)}_k,\dotsc, a'^{(L)}_k)$ and make use of $L$-dimensional ECDQ on the vector $(a'^{(1)}_k,a'^{(2)}_k,\dotsc, a'^{(L)}_k)$.
In general, we will allow $L$ to become large so that,
according to~(\ref{eq:rateecdq}) and the paragraph that follows just
below~(\ref{eq:rateecdq}), the rate loss
$\frac{1}{2}\log_2(2\pi e G_L)$
due to the quantization noise being non-Gaussian
can be made arbitrarily small.
Thus, for large $L$, $E_k$ in (\ref{eq:ratesd1}) can indeed be
approximated as Gaussian noise.
%\vspace{-1.2cm}
%

%%%%%%%%%%%%%%%%%%%%%%%%%%%%%%%%%%%%%%%%%%%%%%%%%%%%%%%%%%%%%%%%%%%%%%%%%%%%
%%%%%%%%%%%%%%%%%%%%%%%%%%%%%%%%%%%%%%%%%%%%%%%%%%%%%%%%%%%%%%%%%%%%%%%%%%%%%

\section{Multiple-Description Coding}
\label{sec:mdchr}
\subsection{MD Delta-Sigma Quantization}\label{sec:mddsq}
\begin{figure*}[th]
\psfrag{ak}{$a_k$}
\psfrag{ak'}{$a'_k$}
\psfrag{akh}{$\hat{a}_k$}
\psfrag{c(z)}{$c'(z)$}
\psfrag{Q}{$Q_L$}
\psfrag{ek}{$e_k$}
\psfrag{ekt}{$\tilde{e}_k$}
\psfrag{xn}{$x_n$}
\psfrag{xn1}{$x_{n'}$}
\psfrag{Description 1}{\small Description 1}
\psfrag{Description 2}{\small Description 2}
\psfrag{Description K}{\small Description K}
\psfrag{h(z)}{$h(z)$}
\psfrag{ha(z)}{$\hspace{-2mm}h_a(z)$}
\psfrag{2}{$2$}
\psfrag{a}{$\alpha$}
\psfrag{b}{$\beta$}
\psfrag{ae}{$\hat{a}_{k,\text{even}}$}
\psfrag{ao}{$\hat{a}_{k,\text{odd}}$}
\psfrag{X1}{$\hat{x}_0$}
\psfrag{X2}{$\hat{x}_1$}
\psfrag{Xc}{$\hat{x}_c$}
\psfrag{hp(z)}{$\hspace{-2mm}h_p(z)$}
\psfrag{QL}{$Q_L$}
\begin{center}
\includegraphics[width=14cm]{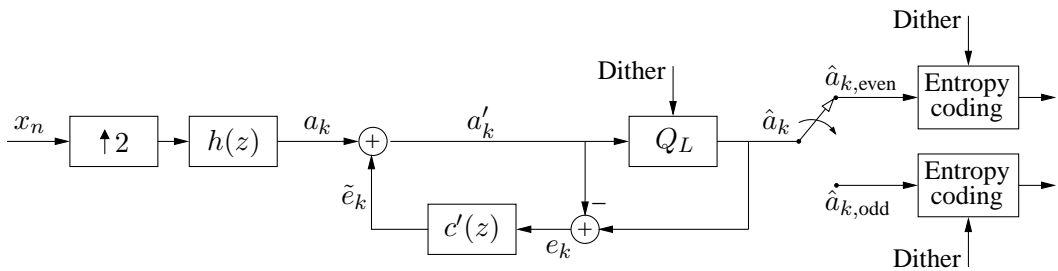}
\caption{Two-channel MD coding based on dithered Delta-Sigma quantization: Encoder.}
\label{fig:deltasigma_md_enc}
\end{center}
%\vspace{-10mm}
\end{figure*}

In this section we show that the sequential dithered Delta-Sigma quantization system,
which is shown in Fig.~\ref{fig:deltasigma},
can be regarded as an MD coding system.
For example, in the case of an oversampling ratio of two,
each input sample leads to two output samples and we have in fact a two-channel
MD coding system as shown in Fig.~\ref{fig:deltasigma_md_enc}.

% As explained in the previous section, we assume that the source is demultiplexed into $L$ parallel
% streams and that an $L$-dimensional lattice quantizer is applied on the set of
% coefficients $(a'^{(1)}_k,\dotsc, a'^{(L)}_k)$.
% However, for illustrational and notational convenience,
% we have only shown a single stream in Fig.~\ref{fig:deltasigma_md_enc}.
% It should also be clear that, as $L$ becomes large, the quantization error
% $E_k = \hat{A}_k - A'_k$
% becomes approximately Gaussian distributed in the sense of the mutual information-rate formula (\ref{eq:ratesd1}).

In the MD scheme of Fig.~\ref{fig:deltasigma_md_enc},
the first description is given by the even outputs of the lattice quantizer and the second description by the odd outputs.
Each description is then entropy-coded separately, conditioned upon its own dither, and transmitted to the decoder.
Notice that although the oversampled signal ${\bs A}$ has memory,
the source part in each description is memoryless,
because we assume ideal interpolation so for a Gaussian source
the even/odd splitting of the samples
corresponds to downsampling by two. 
However, unless the shaped and aliased noise is white and Gaussian, there will be memory in the downsampled signal $\bs{\hat{A}}_{even}$ or $\bs{\hat{A}}_{odd}$. 
We show later that, asymptotically as the vector dimension of the quantizer and the order of the noise-shaping filter approach infinity, the downsampled noise becomes an i.i.d.\ process, and entropy coding can therefore be done sample-by-sample (i.e.\ memorylessly) without loss of optimality. 
By (\ref{eq:ECDQrate}), the sample-by-sample ECDQ rate is given by the (per-sample) block-wise mutual information 
\begin{equation} \label{eq:md_dsq_rate}
R = \frac{1}{L}  I({\bs A'} ; {\bs A'} + {\bs E}).
\end{equation}

%% After quantization, the quantized samples are entropy coded sample-by-sample, i.e.\ in a memoryless fashion (but conditioned upon the dither signal) and then transmitted to the decoder.
% It is important to distinguish between joint versus memoryless (or independent) ECDQ.
% For example, it is customary to assume that the quantizer outputs are encoded jointly whenever
% the source has memory, cf.~\cite{zamir:1995}. If we look at the single-description (SD) setup described
% in Section~\ref{sec:deltasigma}. Then, due to oversampling, the source has memory. If one chooses
% to perform independent entropy coding, then the rate is truly given by the symbol-wise mutual
% information $I(A'_k ; A'_k + E_k)$. However, due to the feedback, this expression is an upper bound
% on the minimum achievable rate. Instead, we should use the directed mutual information~\cite{massey:1990}
% as was done in~\cite{zamir:2007}. Thus, in the SD case, the minimum coding rate is given by the
% directed mutual information. In the MD case, however, each description is memoryless (because of the
% even/odd splitting of the samples, which corresponds to downsampling by two) and the symbol-wise mutual information is therefore identical to the directed mutual information.

\begin{figure}[th]
\psfrag{ak}{$a_k$}
\psfrag{ak'}{$a'_k$}
\psfrag{akh}{$\hat{a}_k$}
\psfrag{c(z)}{$c'(z)$}
\psfrag{Q}{$Q_L$}
\psfrag{ek}{$e_k$}
\psfrag{ekt}{$\tilde{e}_k$}
\psfrag{xn}{$x_n$}
\psfrag{xn1}{$x_{n'}$}
\psfrag{Description 1}{\small Description 1}
\psfrag{Description 2}{\small Description 2}
\psfrag{Description K}{\small Description K}
\psfrag{h(z)}{$h(z)$}
\psfrag{ha(z)}{$\hspace{-2mm}h_a(z)$}
\psfrag{2}{$2$}
\psfrag{a}{$\alpha$}
\psfrag{b}{$\beta$}
\psfrag{ae}{$\hat{a}_{k,\text{even}}$}
\psfrag{ao}{$\hat{a}_{k,\text{odd}}$}
\psfrag{X1}{$\hat{x}_0$}
\psfrag{X2}{$\hat{x}_1$}
\psfrag{Xc}{$\hat{x}_c$}
\psfrag{hp(z)}{$\hspace{-2mm}h_p(z)$}
\psfrag{QL}{$Q_L$}
\begin{center}
\includegraphics[width=8cm]{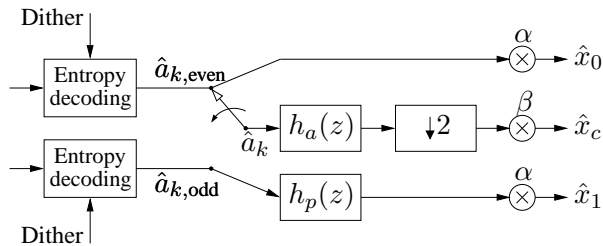}
\caption{Two-channel MD coding based on dithered Delta-Sigma quantization: Decoder.}
\label{fig:deltasigma_md_dec}
\end{center}
%\vspace{-10mm}
\end{figure}

At the decoder, if both descriptions are received, then they are interlaced to form back the oversampled signal $\bs{\hat{A}}$, an anti-aliasing filter $h_a(z)$ (i.e.\ an ideal half-band lowpass filter) is applied and the signal is then downsampled by two and scaled by $\beta$ as shown in Fig.~\ref{fig:deltasigma_md_dec}.
If only the even samples are received, we simply scale the signal by $\alpha$. On the other hand, if only the odd samples are received, we first apply an all-pass filter $h_p(z)$ to correct the phase of the second description and then scale by $\alpha$. The all-pass filter $h_p(z)$ is needed because the upsampling operation performed at the encoder, i.e.\ upsampling by two followed by ideal lowpass filtering (sinc-interpolation), shifts the phase of the odd samples.
The post multipliers $\alpha$ and $\beta$ are described in Section~\ref{sec:dcds}.

%\vspace{-8mm}
%
The distortion due to reconstructing using both descriptions is traditionally called the central distortion $d_c$ and the distortion due to reconstructing using only a single description is called the side distortion $d_s$.

%%%%%%%%%%%%%%%%%%%%%%%%%

\subsection{The Symmetric MD Rate-Distortion Region}\label{sec:mdrdregion}
%The two-channel MD rate-distortion region is completely characterized only in the quadratic
%Gaussian case, i.e.\ the case of memo\-ryless Gaussian sources and MSE fidelity
%criterion~\cite{ozarow:1980,elgamal:1982}.
We will need the solution to the quadratic (memoryless) Gaussian MD problem, as
proven by Ozarow~\cite{ozarow:1980}, in the {\em symmetric} case,
i.e., when both descriptions have the same rate $R$ and the
side distortions are equal.
The set of achievable distortions for description rate $R$
is the union of all distortion pairs $(d_c,d_s)$ satisfying
\begin{equation}\label{eq:ozarowds}
d_s \geq \sigma_X^2 2^{-2R}
\end{equation}
and
\begin{equation}\label{eq:ozarowdc}
d_c \geq \displaystyle\frac{\sigma_X^2 2^{-4R}}{1-(\sqrt{\Pi}-\sqrt{\Delta})^2}
\end{equation}
where $\Pi=(1-d_s/\sigma_X^2)^2$ and $\Delta=d_s^2/\sigma_X^4-2^{-4R}$ and where we require $\Pi\geq\Delta$ to avoid degenerate cases.

Based on the results of~\cite{ozarow:1980}, it was shown in~\cite{vaishampayan:1998c} that
at high resolution, for fixed central-to-side distortion ratio $d_c/d_s$,
the product of the central and side distortions of an optimal two-channel MD scheme
approaches
\begin{equation}\label{eq:distortionproduct1}
d_c d_s \stackrel{\sim}{=} \frac{\sigma_X^4}{4}\frac{1}{1-d_c/d_s}2^{-4R}
\end{equation}
where the approximation $\stackrel{\sim}{=}$ here is in the sense that
the ratio between both sides goes to 1 as $d_s \rightarrow 0$ (or $R \rightarrow \infty$).
If $d_s/d_c\gg 1$,
i.e., at high side-to-central distortion ratio, this simplifies to
\begin{equation}\label{eq:distortionproduct2}
d_c d_s \stackrel{\sim}{=} \frac{\sigma_X^4}{4}2^{-4R}.
\end{equation}

%%%%%%%%%%%%%%%%%%%%%%%%%%%%%%%%%%%%%%%%%%%%%%%%%%%%%%%%%%%%%%%%%%%%%%%%%%%%%%%%%%%%%%%%%%%%%%%%%%%%%

\subsection{Main Theorem}
We now present the main theorem of this work, which basically states that the MD Delta-Sigma
quantization scheme (presented in Section~\ref{sec:mddsq}) can asymptotically achieve the lower bound of
Ozarow's MD distortion region (presented in Section~\ref{sec:mdrdregion}).

\begin{theorem}\label{theo:MD_RDF}
Asymptotically as the noise-shaping filter order $p$ and the vector-quantizer dimension $L$ are going to infinity, % $\rightarrow\infty$,
the entropy rate and the distortion levels
of the dithered Delta-Sigma quantization scheme (of Figs.~\ref{fig:deltasigma_md_enc} and~\ref{fig:deltasigma_md_dec}) with optimum filters and lattice quantizer
achieve the symmetric two-channel MD rate-distortion function
(\ref{eq:ozarowds}) -- (\ref{eq:ozarowdc})
for a memoryless Gaussian source and MSE fidelity criterion,
at any side-to-central distortion ratio $d_s/d_c$
and any resolution.
Furthermore, the optimal infinite-order noise shaping filter is unique, minimum phase, and its magnitude
spectrum $|c(e^{j\omega})|$ is piece-wise flat with a single jump discontinuity at $\omega=\pi/2$.
%
% Rami Z:  Removed the sentence below as it is meaningless at this point.
% Specifically, $|c(e^{j\omega})|^2= \delta^{-1}$ for $|\omega| \leq \pi/2$
% and $|c(e^{j\omega})|^2= \delta$ for $\pi/2< |\omega| < \pi$, where $1 \leq \delta\in \mathbb{R}$.
\end{theorem}

Before presenting the proof of the theorem, we provide in the following sections a series of supporting lemmas. The proof of the theorem can be found in Section~\ref{sec:proofs}.

%%%%%%%%%%%%%%%%%%%%%%%%%%%%%%%%%%%%%%%%%%%%%%%%%%%%%%%%%%%%%%%%%%%%%%%%%%%%%%%%%%%%%%%%%%%%%%%%%%%
%%%%%%%%%%%%%%%%%%%%%%%%%%%%%%%%%%%%%%%%%%%%%%%%%%%%%%%%%%%%%%%%%%%%%%%%%%%%%%%%%%%%%%%%%%%%%%%%%%%

\section{Asymptotic Characterization and Performance Analysis}\label{sec:asympdesc}
In this section we concentrate on the asymptotic case where $p,L\rightarrow \infty$,
i.e.\ infinite noise shaping filter order and infinite vector quantizer dimension.
For analysis purposes, this allows us to assume Gaussian quantization noise
in the system model of Fig.~\ref{fig:deltasigma_model},
with arbitrarily shaped equivalent noise spectrum (\ref{eq:equivalent_spectrum}).
%After gaining some insight from the asymptotic case, we shall turn to
%treat the case of finite $p$ and $L$ in the next section.

%%%%%%%%%%%%%%%%%%

\subsection{Frequency Interpretation of Delta-Sigma Quantization}\label{sec:freq_dsq}
We first give an intuitive frequency interpretation of the proposed Delta-Sigma quantization scheme.
This frequency interpretation reveals that the role of the noise shaping filter is not simply to shape away the quantization noise from the in-band spectrum, as is the case in traditional Delta-Sigma quantization, but rather to delicately control the tradeoff between the in-band noise versus the out-of-band noise, which translates into a tradeoff between the central
and side distortions.
This tradeoff is done while keeping the coding rate fixed, which,
at least at high resolution, is equivalent to keeping the quantizer variance
$\sigma_E^2$ fixed.
See (\ref{eq:ratesdhr}).

%The power spectrum $S_X$ of the i.i.d.\ process $X$ is constant over the complete interval $-\pi$ to $\pi$.
%Now recall that we assume ideal sinc interpolation when resampling.
%As such, since we upsample by a factor of two,
%the power spectrum $S_A$ of the upsampled signal $A$ ranges from $-\pi/2$ to $\pi/2$ as illustrated in Fig.~\ref{fig:spectrum_X_A}.

%The quantization operation, which is based on ECDQ,
%adds a white noise signal $E$ to the over\-sampled signal $A$, after which the feedback
Recall that we, at the central decoder, apply an anti-aliasing filter (ideal lowpass filtering) before downsampling.
Hence, the central distortion is given by the energy of the quantization noise that falls within the
in-band spectrum.
The inclusion of a noise shaping filter at the encoder makes it possible to shape away the quantization noise from the in-band spectrum and thereby reduce the central distortion.
By increasing the order of the noise shaping filter it is possible to reduce the central distortion accordingly.

\begin{figure}[th]
\psfrag{SE}{\small $S_E$}
\psfrag{pi}{\small $\pi$}
\psfrag{-pi}{\small $-\pi$}
\psfrag{SCE}{\small$\!\!|c(e^{j\omega})|^2\sigma_E^2$}
\psfrag{pi2}{\small $\pi/2$}
\psfrag{-pi2}{\small $-\pi/2$}
\psfrag{w}{\small $\omega$}
\psfrag{0}{\small $0$}
\psfrag{s}{\raisebox{0.5mm}{$\scriptstyle \sigma_E^2$}}
\psfrag{ds}{\raisebox{1mm}{$\scriptstyle \sigma_E^2\delta$}}
\psfrag{ids}{\raisebox{0mm}{$\scriptstyle \sigma_E^2/\delta$}}
\begin{center}
\subfigure[Spectrum of $E$]{\includegraphics[width=5.6cm]{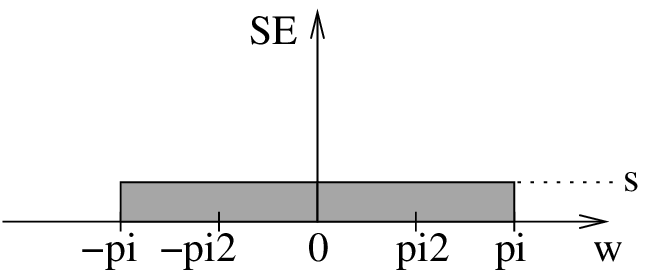}}\par
\hspace{3.5mm}\subfigure[Spectrum of shaped $E$]{\includegraphics[width=6cm]{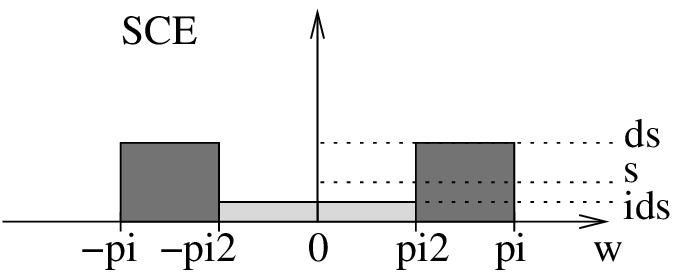}\label{fig:spectrum_CE}}
\caption{The power spectrum of (a) the quantization noise (b) the shaped quantization noise. In (b) the energy of the lowpass noise spectrum (the bright region) corresponds to the central distortion and the energy of the full spectrum corresponds to the side distortion.}
\label{fig:spectrum_E}
\end{center}
%\vspace{-10mm}
\end{figure}

It is also interesting to understand what influences the side distortion. The side descriptions are constructed by using either all odd samples or all even samples of the output $A$. Hence, we effectively downsample $A$ by a factor of two.
It is important to see that this downsampling process takes place without first applying an anti-aliasing filter. Thus, aliasing is inevitable. It follows, that not only the noise which falls within the in-band spectrum contributes to the side distortion but also the noise that falls outside the in-band spectrum (i.e.\ the out-of-band noise) affects the distortion. Since, in traditional Delta-Sigma quantization, the noise is shaped away from the in-band spectrum as efficiently as possible, the out-of-band noise is likely to be the dominating contributor to the side distortion. We have illustrated this in Fig.~\ref{fig:spectrum_E}.
%\vspace{-5mm}
%

It should now be clear that, in two-channel MD Delta-Sigma quantization, the role of the noise shaping filter is to trade off the in-band noise versus the out-of-band noise.
In particular, in the asymptotic case where the order of the noise shaping filter goes to infinity, it is possible to construct a brick-wall filter which has a squared magnitude spectrum of $1/\delta$ in the passband (i.e.\ for $|\omega| \leq \pi/2$) and of $\delta$ in the stopband (i.e.\ for $\pi/2 < |\omega| < \pi$). In this case, the central distortion is proportional to $1/\delta$ whereas the side distortion is proportional to $1/\delta + \delta$.
This situation, which is illustrated in Fig.~\ref{fig:spectrum_CE}, will be discussed in more detail in the next section.%Section~\ref{sec:distproduct}.

%%%%%%%%%%%%%%%%%%%%%%%%%%%%%%%%%%%%

\subsection{Achieving the MD Distortion Product at High Resolution}\label{sec:distproduct}
It is possible to take advantage of the frequency interpretation given in Section~\ref{sec:freq_dsq}
in order to show that the optimum central-side distortion product
at high-resolution~(\ref{eq:distortionproduct1})
can be achieved by Delta-Sigma quantization.
We later extend this result and show that with suitable post-multipliers at the decoders,
optimum performance is achieved at {\em any} resolution.

\begin{lemma}\label{lem:distproduct}
At high resolution and asymptotically as $p,L\rightarrow\infty$,
the distortion product given by~(\ref{eq:distortionproduct1})
is achievable by Delta-Sigma quantization.
\end{lemma}%

\begin{IEEEproof}
The central distortion is equal to the total energy $\sigma_E^2P_{d_c}$ of the in-band noise spectrum where
\begin{equation}\label{eq:Pdc}
P_{d_c}=\frac{1}{2\pi}\int_{-\pi/2}^{\pi/2} |c(e^{j\omega})|^2d\omega.
\end{equation}
%and recall that the oversampled source has variance $\frac{1}{2\pi}\int_{-\pi}^{\pi}S_A(\omega)\, d\omega=\sigma_X^2$ where $S_A(\omega)=2\sigma_X^2$ for $|\omega|\leq\frac{\pi}2$ and $S_A(\omega)=0$ otherwise. 

The side distortion is equal to the energy $\sigma_E^2P_{d_s}$ of the
in-band noise spectrum of the side descriptions which contains aliasing
due to the subsampling process. Since we downsample by two we have
\begin{equation}\label{eq:Pds}
P_{d_s}=\frac{1}{4\pi}\int_{-\pi}^{\pi}|c(e^{j\omega/2})|^2 + |c(e^{j(\omega/2+\pi)})|^2d\omega.
\end{equation}
%and the source variance remains at $\sigma_X^2$ after downsampling by two.

Let us shape the noise spectrum as illustrated in Fig.~\ref{fig:spectrum_CE}. Thus, we let $|c(e^{j\omega})|^2 = 1/\delta$ for $|\omega|\leq \pi/2$ and $|c(e^{j\omega})|^2 = \delta$ for $\pi/2<|\omega|<\pi$ where $0<\delta\in \mathbb{R}$.
It follows from~(\ref{eq:Pds}) that, for any $\delta>0$, $d_s = \frac{1}{2}\sigma_E^2(\delta+\delta^{-1})$ and from~(\ref{eq:Pdc}) we see that $d_c=\frac{1}{2}\sigma_E^2/\delta$ which yields the distortion product 
\begin{equation}\label{eq:dist_product_dcds}
d_cd_s = \frac{\delta+\delta^{-1}}{4\delta}\sigma_E^4.
\end{equation}
From~(\ref{eq:ratesdhr}) we know that at high resolution
$R\approx \frac{1}{2}\log_2(\sigma_X^2/\sigma_E^2)$ (where $\approx$ is in the sense that the difference goes to zero
as $R \rightarrow \infty$),
so that 
\begin{equation}\label{eq:sigmaE4}
\sigma_E^4 \stackrel{\sim}{=} \sigma_X^4 2^{-4R}
\end{equation}
(where $\stackrel{\sim}{=}$ is in the sense that the ratio goes to one
as $R \rightarrow \infty$).
Finally, since $d_c/d_s = \delta^{-1}/(\delta+\delta^{-1})$ it follows that
\begin{equation}\label{eq:dist_ratio}
\frac{1}{1-d_c/d_s} = \frac{\delta+\delta^{-1}}{\delta}
\end{equation}
and the lemma is proved by inserting~(\ref{eq:sigmaE4}) and~(\ref{eq:dist_ratio}) into~(\ref{eq:dist_product_dcds}) and comparing the resulting expression to~(\ref{eq:distortionproduct1}).
\end{IEEEproof}

%%%%%%%%%%%%%%%%%%%%%%%%%%%%%%%%%%%%%%%%%%%%%%%%%%%%%%%%%%%%%%%%%%%%%%

\subsection{Optimum Performance for General Resolution}\label{sec:dcds}
%% \subsection{Central and Side Distortions}\label{sec:dcds}
In this section we extend the optimality result of Section~\ref{sec:distproduct} above,
% we assess the side distortion $d_s$ and central distortion $d_c$ of the two-channel Delta-Sigma quantization scheme
and show that the two-channel Delta-Sigma quantization scheme achieves
the symmetric quadratic Gaussian rate-distortion function at any resolution.

Let $U_i$ denote the reconstructions before the side post multipliers so that $\hat{X}_i = \alpha U_i, i=0,1$, and let $\mathbb{E}$ denote the expectation operator. It can then be shown that $\mathbb{E}XU_i = \sigma_X^2$ and $\mathbb{E}U_i^2 = \sigma_X^2 + \sigma_E^2(\delta+\delta^{-1})/2$. Moreover, let $U$ denote the reconstruction before the central multiplier $\beta$. Then $\mathbb{E}U^2 = \sigma_X^2 + \sigma_E^2\delta^{-1}/2$. Finally, let the post multipliers be given by
\begin{equation*}
\alpha = \frac{\sigma_X^2}{\sigma_X^2 + \sigma_E^2(\delta+\delta^{-1})/2}
\end{equation*}
and
\begin{equation*}
\beta = \frac{\sigma_X^2}{\sigma_X^2 + \sigma_E^2\delta^{-1}/2}.
\end{equation*}
It follows that the side distortion is given by
\begin{align} \notag
d_s &= \mathbb{E}(\hat{X}_i-X)^2  \\ \notag
    &= \mathbb{E}(\alpha U_i - X)^2 \\ \label{eq:beforesidepost}
    &= \sigma_X^2 - 2\alpha\sigma_X^2 + \alpha^2(\sigma_X^2+\sigma_E^2(\delta+\delta^{-1})/2) \\ \label{eq:sidedistpost}
&= \frac{\sigma_X^2\sigma_E^2(\delta+\delta^{-1})}{2\sigma_X^2 + \sigma_E^2(\delta+\delta^{-1})}.
\end{align}
Similarly, let $\hat{X}_c = \beta U$ so that the central distortion is given by
\begin{align} \notag
d_c &= \mathbb{E}(\hat{X}_{c}-X)^2 \\ \notag
&= \mathbb{E}(\beta U - X)^2 \\ \label{eq:beforecentralpost}
&= \sigma_X^2 + \beta^2(\sigma_X^2 + \sigma_E^2\delta^{-1}/2) - 2\beta\sigma_X^2 \\ \label{eq:centralpost}
&= \frac{\sigma_X^2\sigma_E^2\delta^{-1}}{2\sigma_X^2+\sigma_E^2\delta^{-1}}.
\end{align}

\begin{lemma}\label{lem:optqgrdf}
For a given description rate $R$ and
asymptotically as $p,L\rightarrow\infty$
(i.e., assuming Gaussian quantization noise and equivalent noise spectrum as in
Fig.~\ref{fig:spectrum_CE}),
the side distortion given by~(\ref{eq:sidedistpost}) and the central distortion given
by~(\ref{eq:centralpost}) achieve the lower bound (\ref{eq:ozarowdc})
of Ozarow's symmetric MD distortion region.
\end{lemma}

\begin{IEEEproof}
Recall from Section~\ref{sec:deltasigma},
that the rate of memoryless-ECDQ
(assuming that the entropy coding is conditioned upon the dither signal
and that the dither signal is known at the decoder)
is equal to the
mutual information between the input and the output of an additive noise channel
(\ref{eq:md_dsq_rate}).
For large $L$, this mutual information can be calculated as if
the additive noise $E_k$ was approximately Gaussian distributed.
It thus follows from~(\ref{eq:ratesd1}) and~(\ref{eq:ratesd2}) that as $L \rightarrow \infty$
the description rate becomes
\begin{align} \notag
R&=I(A'_k ; \hat{A}_k) \\ \notag
&= h(\hat{A}_k) - h(E_k) \\ \notag
&= \frac{1}{2}\log_2(2\pi e (\sigma_X^2+\sigma_E^2(\delta+\delta^{-1})/2)) - \frac{1}{2}\log_2(2\pi e\sigma_E^2) \\ \label{eq:descrate}
&= \frac{1}2\log_2\left(\frac{\sigma_X^2 + \sigma_E^2(\delta+\delta^{-1})/2}{\sigma_E^2}\right) .
\end{align}
We can rewrite~(\ref{eq:descrate}) as
\begin{equation}\label{eq:ratemisc}
2^{-4R}=\frac{4\delta^2\sigma_E^4}{(2\sigma_X^2\delta + \sigma_E^2\delta^2 + \sigma_E^2)^2}.
\end{equation}
By use of~(\ref{eq:sidedistpost}) and~(\ref{eq:ratemisc}) we then get
\begin{equation*}
\begin{split}
\Delta = \frac{\sigma_E^4(\delta^4-2\delta^2+1)}{(2\sigma_X^2\delta + \sigma_E^2\delta^2 + \sigma_E^2)^2}
\end{split}
\end{equation*}
and
\begin{equation*}
\Pi = \frac{4\delta^2\sigma_X^4}{(2\sigma_X^2\delta + \sigma_E^2\delta^2 + \sigma_E^2)^2}
\end{equation*}
so that
\begin{equation}\label{eq:PiDelta}
1-(\sqrt{\Pi}-\sqrt{\Delta})^2 = \frac{4\sigma_E^2\delta^2(2\sigma_X^2\delta+\sigma_E^2)}{(2\sigma_X^2\delta + \sigma_E^2\delta^2 + \sigma_E^2)^2}.
\end{equation}
Finally, inserting~(\ref{eq:PiDelta}) in~(\ref{eq:ozarowdc}) leads to
\begin{equation*}
\displaystyle\frac{\sigma_X^2 2^{-4R}}{1-(\sqrt{\Pi}-\sqrt{\Delta})^2}=
\frac{\sigma_X^2\sigma_E^2}{2\sigma_X^2\delta + \sigma_E^2}
\end{equation*}
which is identical to~(\ref{eq:centralpost}) and therefore proves the lemma.
\end{IEEEproof}

%%%%%%%%%%%%%%%%%%%%%%%%%%%%%%%

\subsection{Relation to Ozarow's Double Branch Test Channel}\label{sec:dsq_ozarow}
\begin{figure}[th]
\psfrag{N0}{\raisebox{1mm}{$N_0$}}
\psfrag{N1}{\raisebox{-1mm}{$N_1$}}
\psfrag{X}{$X$}
\psfrag{X0}{$\hat{X}_0$}
\psfrag{X1}{$\hat{X}_1$}
\psfrag{Xc}{$\hat{X}_c$}
\psfrag{a0}{\raisebox{1mm}{$\alpha_0$}}
\psfrag{a1}{\raisebox{1mm}{$\alpha_1$}}
\psfrag{b0}{$\beta_0$}
\psfrag{b1}{$\beta_1$}
\psfrag{U0}{$U_0$}
\psfrag{U1}{$U_1$}
\begin{center}
\includegraphics[width=5.5cm]{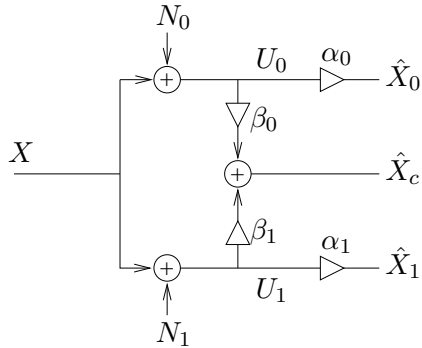}
\caption{The MD optimum test channel of Ozarow~\cite{ozarow:1980}. At high resolution $\alpha_i=1$ and $\beta_i=1/2, i=0,1$ so that $\hat{X}_0=U_0, \hat{X}_1=U_1$ and $\hat{X}_c=\frac{1}{2}(\hat{X}_0+\hat{X}_1)$.}
\label{fig:testchannel}
\end{center}
%\vspace{-1cm}
\end{figure}

Let us now revisit Ozarow's double branch test channel as shown in Fig.~\ref{fig:testchannel}. In this model the noise pair $(N_0,N_1)$ is \emph{negatively} correlated (except from the case of no-excess marginal rates, in which case the noises are independent).
Notice that this is in line with the above observations, since the highpass nature of the noise shaping filter causes adjacent noise samples to be negatively correlated. The more negatively correlated they are, the greater is the ratio of side distortion to central distortion. Furthermore, at high resolution, the ``filters'' ($\alpha_i$ and $\beta_i, i=0,1$) in Ozarow's test channel become trivial and the central reconstruction is simply given by the average of the two side channels. This averaging operation can be seen as a lowpass filtering operation, which leaves the signal (since it is lowpass) and the in-band noise intact but removes the out-of-band noise.

More formally, for the symmetric case (where $\sigma_{N}^2=\sigma_{N_i}^2,i=0,1$ and $\rho$ is the correlation coefficient of the noises), we have the following high-resolution relationships between $(\rho,\sigma_N^2)$ of Ozarow's test channel and $(\delta,\sigma_E^2)$ of the proposed Delta-Sigma quantization scheme.
\begin{lemma}
At high-resolution conditions, we have
\begin{equation}\label{eq:sigma_EN}
\sigma_E^2 = \sigma_N^2 \sqrt{1-\rho^2}
\end{equation}
and
\begin{equation}\label{eq:delta_rho}
\delta = \frac{ \sqrt{1-\rho} }{ \sqrt{1+\rho} }.
\end{equation}
\end{lemma}

\begin{IEEEproof}
From~\cite{zamir:1999,zamir:2000} it follows that Ozarow's sum rate $R_0+R_1$ satifies
\begin{align} \notag
R_0 + R_1 &\geq I(X; X+N_0) + I(X; X + N_1)  \\ \notag
&\quad+ I(X + N_0 ; X + N_1) \\ \notag
&= I(X; X+N_0) + I(X; X + N_1) + I(N_0;N_1) \\ \notag
 &= I(X; X+N_0) + I(X; X + N_1) \\ \notag
&\quad  + \frac{1}{2}\log_2\left(\frac{1}{1-\rho^2}\right) \\ \notag
&= h(X+N_0) - h(X+N_0|X) + h(X+N_1)  \\ \notag 
&\quad- h(X+N_1|X) - \log_2(\sqrt{1-\rho^2}) \\  \label{eq:sumrate}
&= \log_2\left(\frac{\sigma_X^2 + \sigma_N^2}{\sigma_N^2}\right) - \log_2\left(\sqrt{1-\rho^2}\right)
\end{align}
where the last equality follows since the noises have equal variances. 
By equating~(\ref{eq:sumrate})
to~(\ref{eq:descrate}), i.e.\ $2R=\log_2(\frac{\sigma_X^2 + \sigma_E^2(\delta + \delta^{-1})/2}{\sigma_E^2})$ and solving for $\sigma_E^2$ we obtain
\begin{equation}\label{eq:sigma_EN_gen}
\sigma_E^2 = \frac{2\sigma_X^2\sigma_N^2\sqrt{1-\rho^2}}{ 2(\sigma_X^2 + \sigma_N^2) - (\delta + \delta^{-1})\sigma_N^2\sqrt{1-\rho^2}}.
\end{equation}
If $\sigma_N^2\ll\sigma_X^2$, this reduces to $\sigma_E^2\approx \sigma_N^2\sqrt{1-\rho^2}$ and we obtain~(\ref{eq:sigma_EN}).

The MMSE when estimating $X$ from two jointly Gaussian noisy observations $U_i=X+N_i, i=0,1$ (where the Gaussian noises have equal variances), is given by
\begin{equation}\label{eq:mmse}
\text{MMSE} = \frac{\sigma_N^2(1+\rho)}{\sigma_N^2(1+\rho) + 2}.
\end{equation}
Thus, the central distortion of Ozarow's test channel is given by~(\ref{eq:mmse}), which we equate to the central distortion~(\ref{eq:centralpost}),
%$d_c= \frac{\sigma_X^2\sigma_E^2}{2\sigma_X^2\delta + \sigma_E^2}$
solve for $\delta$ and insert $\sigma_E^2$ from (\ref{eq:sigma_EN_gen}), that is
\begin{equation}\label{eq:delta_gen}
\begin{split}
\delta &= \frac{\sigma_X^2(\sigma_N^2(1+\rho)+2)}{2(\sigma_X^2+\sigma_N^2) - (\delta + \delta^{-1})\sigma_N^2\sqrt{1-\rho^2}}\cdot \frac{\sqrt{1-\rho}}{\sqrt{1+\rho}} \\
&\quad- \frac{\sigma_N^2\sqrt{1-\rho^2}}{2(\sigma_X^2+\sigma_N^2) - (\delta + \delta^{-1})\sigma_N^2\sqrt{1-\rho^2}}.
\end{split}
\end{equation}
Once again, letting $\sigma_N^2\ll \sigma_X^2$ it follows that $\delta\approx \frac{\sqrt{1-\rho}}{\sqrt{1+\rho}}$, which yields~(\ref{eq:delta_rho}) and thereby proves the lemma.
\end{IEEEproof}

\begin{remark}
The relationship between Ozarow's test-channel and the Delta-Sigma quantization scheme at general resolution is provided by~(\ref{eq:sigma_EN_gen}) and~(\ref{eq:delta_gen}).
\end{remark}

\section{Proof of Theorem~\ref{theo:MD_RDF}}\label{sec:proofs}
We are now in a position to wrap up the proof of Theorem~\ref{theo:MD_RDF}.
Lemma~\ref{lem:optqgrdf} actually shows that it is possible to achieve the quadratic
Gaussian rate-distortion function if we replace the ECDQ by a Gaussian noise,
and the equivalent noise spectrum (\ref{eq:equivalent_spectrum})
by a brick wall spectrum.  This can be viewed as setting the lattice
quantizer dimension $L$ and the feedback filter order $p$ to be {\em equal
to infinity}.
Thus, what is still missing is the characterization of the limit
behavior of the coding rate as $L, p\rightarrow\infty$,
and the distortion as $p\rightarrow\infty$. 

\subsection{Distortion loss}
We first present Lemma~\ref{lem:dcpdc} (with a proof in the appendix) which describes the central and the side distortions at general resolution when using an arbitrary noise-shaping filter $c(e^{j\omega})$. 
\begin{lemma}\label{lem:dcpdc}
For any given $p$th-order noise-shaping filter $c(e^{j\omega})$ and optimal multipliers ($\alpha$ and $\beta$), the central distortion is given by
\begin{equation}\label{eq:dcPdc}
d_c = \frac{\sigma_X^2 \sigma_E^2P_{d_c}}{\sigma_X^2 + \sigma_E^2P_{d_c}}
\end{equation}
and the side distortion is given by
\begin{equation}\label{eq:dsPds}
d_s = \frac{\sigma_X^2 \sigma_E^2P_{d_s}}{\sigma_X^2 + \sigma_E^2P_{d_s}}
\end{equation}
where 
\begin{equation*}
P_{d_c} = \frac{1}{2\pi}\int_{|\omega|\leq \frac{\pi}2} |c(e^{j\omega})|^2 d\omega = \frac{1}{2}\sum_{i=0}^p\sum_{j=0}^{p}\sinc\left(\frac{i-j}{2}\right)c_ic_j
\end{equation*}
and
\begin{equation*}
P_{d_s} = \frac{1}{2\pi}\int_{|\omega|\leq \pi} |c(e^{j\omega})|^2 d\omega = \sum_{j=0}^{p}c_j^2.
\end{equation*}
\el
\end{lemma}

The noise shaping filter used in the proof of Lemma~\ref{lem:optqgrdf} to show achievability of the quadratic Gaussian rate-distortion function is of infinite order, and it satisfies
\begin{equation}\label{eq:minphaseprop}
\frac{1}{2\pi}\int_{-\pi}^{\pi}\log_2|c(e^{j\omega})|^2d\omega = 0.
\end{equation}
It follows that the area under $\log_2|c(e^{j\omega})|^2$ is equally distributed above and below the 0 dB line, which is a unique property of minimum-phase filters~\cite{markel:1976}. In fact, the following Lemma proves that, in order for $c(z)$ to be optimal, it must be of infinite-order and minimum phase (see the appendix for the proof). This means that the optimum noise shaping filter is unique.

\begin{lemma}\label{lem:opt_uniq}
In order to achieve the quadratic Gaussian rate-distortion function, it is required that the noise shaping filter $c(z)$ is of infinite order, 
minimum-phase, and have a piece-wise flat power spectrum of power $\delta^{-1}$ in the lowpass band (i.e.\ for $|\omega| < \pi/2$)
% or equivalently for $|f|\leq 1/2$) 
and of power $\delta$ in the highpass band (i.e.\ for $\pi/2< |\omega| < \pi$)
% or equivalently $1/4 < |f| < 1/2$) 
where $1 \leq \delta\in \mathbb{R}$.
\el
\end{lemma}

\begin{remark}
It can be observed from~(\ref{eq:delta_rho}) that (at high resolution) choosing $\delta$ in the range $0<\delta<1$ corresponds to having positively correlated noises in Ozarow's test-channel. Moreover, the central distortion is proportional to $1/\delta$ whereas the side distortion as well as the description rate are both proportional to the sum $\delta + 1/\delta$. Thus, while it is not a system requirement to choose $\delta \geq 1$, it is a better choice performance wise.
\end{remark}

We now assess the distortion loss due to using a finite order noise-shaping filter.
Let $S^{opt}_\epsilon(\omega)=\sigma_E^2|c^{opt}(e^{j\omega})|^2$ denote the power spectrum of the shaped noise when using the ideal infinite-order noise shaping filter $c^{opt}(e^{j\omega})$, which is optimal and unique as proven by Lemma~\ref{lem:opt_uniq}.
Thus, $|c^{opt}(e^{j\omega})|^2$ is piece-wise flat with a jump discontinuity at $\omega=\pi/2$, cf.\ Fig.~\ref{fig:spectrum_CE}. For such a function, point-wise convergence of the Fourier coefficients cannot be guaranteed. However, we do have convergence in the mean square sense~\cite{bachman:2000}. Specifically, let 
$S_\epsilon^{(p)}(\omega)$ denote the $p$th order Fourier approximation to $S_\epsilon^{opt}(\omega)$. Then~\cite{bachman:2000}
\begin{equation}\label{eq:gibbs}
\lim_{p\rightarrow\infty}\frac{1}{2\pi}\int_{|\omega|\leq \pi} \left| S_\epsilon^{opt}(\omega) - S_\epsilon^{(p)}(\omega)\right|^2d\omega = 0
\end{equation}
which asserts that the limit for $p\rightarrow\infty$ exists. In addition, it can be shown that the error (MSE) of the $p$th order Fourier approximation of this step function is of the order $\mathcal{O}(1/p)$~\cite{vetterli:2001}.
It follows that for any $p$ we have
\begin{equation}\label{eq:dcPdc2}
\begin{split}
d_c &= \frac{\sigma_X^2 \sigma_E^2(P_{d_c} + \mathcal{O}(1/p))}{\sigma_X^2 + \sigma_E^2(P_{d_c} + \mathcal{O}(1/p))},\\
d_s &= \frac{\sigma_X^2 \sigma_E^2(P_{d_s} + \mathcal{O}(1/p))}{\sigma_X^2 + \sigma_E^2(P_{d_s} + \mathcal{O}(1/p))}.
\end{split}
\end{equation}
and the desired continuity in the limit $p\rightarrow\infty$ is established.

\subsection{Coding rate loss}
By similar arguments as leading to~(\ref{eq:dcPdc2}), for a finite $p$, the variance of $\hat{A}_k$ is given by
\begin{align}
Var(\hat{A}_k) &= \mathbb{E}[ (A'_k + E_k)^2] \\ \label{eq:var_increase}
&= \sigma_X^2 + 
\sigma_E^2(P_{ds} + \mathcal{O}(1/p ))
\end{align}
Moreover, the coding rate~(\ref{eq:ratesd1}) when using memoryless entropy coding is given by 
\begin{align}\notag %\label{eq:finiteLR}
R &= I(A'_k ; \hat{A}_k)= h(\hat{A}_k) - h(E_k) \\ \label{eq:div_noise}
  &= h(\hat{A}_k) - h(E_k^*) + \frac{1}{2}\log_2( G_L 2\pi e) \\ \label{eq:div_noise1}
  &= h(\hat{A}_k) - h(E_k^*) + \mathcal{O}(\log_2(L)/L)
\end{align}
where $h(\hat{A}_k)$ is an increasing function in $Var(\hat{A}_k)$ and 
$E_k^*$ denotes a Gaussian variable (process) with the same variance (spectrum) as $E_k$. Eq.~(\ref{eq:div_noise}) follows from the discussion after~(\ref{eq:rateecdq}) where it may be noticed that the term $\frac{1}{2}\log_2( G_L 2\pi e)$ describes the divergence of the quantization noise from Gaussianity; see also~\cite{zamir:1996b}.
This divergence term corresponds to an excess rate due to using a finite dimensional lattice quantizer and may be upper bounded by $\mathcal{O}(\log_2(L)/L)$ when optimal $L$-dimensional lattice quantizers are used, see~\cite{zamir:1996} for details.
Thus, if we keep $\sigma_E^2$ fixed, then the coding rate is increased due to the $\mathcal{O}(1/p)$ variance increase given in~(\ref{eq:var_increase}) and due to the excess term $\mathcal{O}(\log_2(L)/L)$ in~(\ref{eq:div_noise1}). 
These rate penalties vanish as $p,L\rightarrow\infty$ and the desired convergence in coding rate is proved. 

In order to complete the proof of the theorem, we need to show that an optimal monic minimum phase filter always exists for any ratio of $P_{d_s}/P_{d_c}$.
Towards that end, we keep the post multipliers fixed and define $J =  \lambda_c P_{d_c} + \lambda_s P_{d_s}$ as the cost function to be minimized by the $p$th-order noise-shaping filter. Notice that if we let $\lambda_s=0$ we are only concerned about minimizing the noise power than falls in the in-band region. Thus, we are aiming at minimizing the central distortion. On the other hand, letting $\lambda_c \ll \lambda_s$ gives priority to the side distortion since the total noise power is minimized.
Let $c_0=1$ and $\bs{c}=(c_1,\dotsc,c_p)$ be the filter coefficients. Moreover, let $\bs{g}$ be the $p$-vector with elements $g_i=\sinc(i/2), i=1,\dotsc,p$, and let $\bs{G}$ be the $p\times p$ autocorrelation matrix with elements $G_{i,j}=\sinc((i-j)/2)$, where $i,j\in \{1,\dotsc,p\}$. With this it follows that
\begin{equation*}
P_{d_s}=\sum_{i=0}^{p} c_i^2 = (1 + \bs{c}^T\bs{c})
\end{equation*}
and 
\begin{equation*}
\begin{split}
P_{d_c}&=\frac{1}{2}\sum_{i=0}^{p}\sum_{j=0}^{p}\sinc\left(\frac{i-j}{2}\right)c_ic_j  \\
&= \frac{1}{2}(1 + 2\sum_{i=1}^{p}\sinc(i/2)c_i + \sum_{i=1}^{p}\sum_{j=1}^{p}\sinc((i-j)/2)c_ic_j) \\
&= \frac{1}{2}(1 + 2\bs{c}^T\bs{g} + \bs{c}^T\bs{G}\bs{c})
\end{split}
\end{equation*}
so that
\begin{equation*}
\begin{split}
\lambda_c P_{d_c} \!&+\! \lambda_s P_{d_s} = \frac{1}{2}\left( \lambda_c(1+2\bs{c}^T\bs{g} + \bs{c}^T\bs{G}\bs{c}) + 2\lambda_s(1+ \bs{c}^T\bs{c})\right) \\
&= \frac{1}{2}\left( \lambda_c + 2\lambda_s + 2\lambda_c\bs{c}^T\bs{g} + \bs{c}^T(\lambda_c\bs{G}+2\lambda_s\bs{I})\bs{c}\right).
\end{split}
\end{equation*}
The optimal filter coefficients are found by solving the differential equation $\frac{\partial}{\partial \bs{c}}(\lambda_c P_{d_c} + \lambda_s P_{d_s})= \bs{0}$, that is
\begin{equation}\label{eq:optbsum}
\bs{c}=-(\bs{G}+2\frac{\lambda_s}{\lambda_c}\bs{I})^{-1}\bs{g}
\end{equation}
where $\bs{I}$ is the $p\times p$ identity matrix. Notice that $\bs{G}+2\frac{\lambda_s}{\lambda_c}\bs{I}$ is a symmetric and full rank matrix and~(\ref{eq:optbsum}) therefore defines a well-posed problem. The solution to~(\ref{eq:optbsum}) can be found by the Yule-Walker method, which yields a unique minimum-phase filter~\cite{makhoul:1975}. 
As $p\rightarrow\infty$, the autocorrelation sequence of the impulse response of the obtained filter $c(z)$ becomes identical to the \emph{ideal} autocorrelation sequence whose Fourier transform describes the optimum shaped noise spectrum~\cite{makhoul:1975}.
Thus, the resulting spectrum of the shaped noise becomes identical to the optimum spectrum. This proves the theorem.

% \begin{remark}
% The filter coefficients given by Lemma~\ref{lem:msefilter} are in fact equivalent to those presented in~\cite{tewksbury:1978} where the in-band noise of a noise shaping coder is minimized in the frequency domain.
% \end{remark}

%%%%%%%%%%%%%%%%%%%%%%%%%%%%%%%%%%%%%%%%%%%%%%%%%%%%%%%%%%%%%%%%%%%%%%%%%%%%%%%%%%%%%%%%%%%%%%%%%%%%%
%\vspace{-2mm}
\section{Universality of Dithered Delta-Sigma Quantization}
\label{sec:universal}
%\vspace{-1mm}
In this section we discuss the universality of the proposed scheme at high resolution. 
First, notice that the central and side
distortions depend only upon the second-order statistics of the source and
the quantization noise, i.e.\ $\sigma_X^2$ and $\sigma_E^2$, and as such
not on the Gaussianity of the source.
Second, independent of the source distribution, the distribution of the
quantization noise becomes approximately Gaussian distributed (in the
divergence sense) in the limit of high vector quantizer dimension $L$.
Finally, the ECDQ is allowed to encode each description according to its
entropy. Thus, the coding rate is equal to the mutual information
(\ref{eq:md_dsq_rate})
of the source over the Gaussian test channel.
For memoryless sources of equal variances, this coding rate is upper
bounded by that of the Gaussian source.
Moreover, Zamir proved in~\cite{zamir:1999} that Ozarow's test channel
becomes asymptotically optimal in the limit of high resolution for any
i.i.d. source provided it has a finite differential entropy. Thus, since
the dithered Delta-Sigma quantization scheme resembles Ozarow's test
channel in the limit as $p,L\rightarrow\infty$,
we deduce that the proposed scheme becomes asymptotically
optimal for general i.i.d.\ sources with finite differential entropy.

A delicate point to note, though, is that
due to the \sinc\ interpolation, the odd samples might not be i.i.d.\
and joint entropy coding within the packet is necessary in order to be
optimal.   Specifically,
with joint entropy coding the rate is given by the \emph{directed} mutual
information formula (\ref{eq:directed})
applied to the sub-sampled source $\hat{A}_{k,odd}$.
The resulting rate for the odd packet is
$\bar{h}(A_{k,odd}) - h(E_k)$,
which (at high resolution) is
$\approx h(X) - \frac{1}{2}\log_2(2 \pi e \sigma_E^2)$,
as desired \cite{zamir:1999}.

If we have a \emph{source with memory}, and we allow joint entropy
coding within each of the two packets,
then a similar derivation shows that we would achieve rate
$R \approx \bar{h}(X) - \frac{1}{2}\log_2(2 \pi e \sigma_E^2)$
in each packet. This rate is the mutual information rate of the source
over the Gaussian test channel. Since Ozarow's test channel is
asymptotically optimal in the limit of high resolution for any stationary
source with finite differential entropy rate, \cite{zamir:2000}, it
follows that the proposed scheme is asymptotically optimal for such sources
as well. It is also interesting to note that
we recently proposed to combine noise-shaping with prediction when encoding sources with memory~\cite{kochman:2008}. In the Gaussian case, this makes it possible to remove the source redundancy without requiring entropy coders with memory.

\section{Extension to $K>2$ descriptions}\label{sec:extensionK}
We end this paper by presenting a straight-forward extension of the proposed design to $K$ descriptions, though without any claim of optimality.
The basic idea is to change the oversampling ratio from two to $K$ and then decide which output samples should make up a description.\footnote{Notice that even fractional oversampling ratios  can be used in which case we might also have aliasing of the source spectrum.}
When dealing with $K$ descriptions, $2^K - 1$ distinct subsets of descriptions can be created. Thus, the design of the decoders is generally more complex for greater $K$. For example, if two out three descriptions are received, aliasing is unavoidable (as was the case for $K=2$ descriptions). Moreover, due to the fractional (non-uniform) downsampling process, the simple brick-wall lowpass filter operation is not necessarily the optimal reconstruction rule. In fact, the optimal reconstruction rule depends not only upon the number of received descriptions but (generally) also upon which descriptions are received.
However, in this section we will restrict attention to cases leading to uniform sampling.\footnote{We suspect that results from non-uniform sampling or non-uniform filterbank theory will prove advantageous for constructing the optimal decoders in the most general situation. However, this is a topic of future research.}
Thus, the design of the decoders is simplified.

We use the previously presented Delta-Sigma quantization scheme (of Figs.~\ref{fig:deltasigma_md_enc} and~\ref{fig:deltasigma_md_dec}) but oversample now by $K$ instead of two. More specifically, let us assume that $K=4$ and that every fourth sample make up a description. We notice that the extension to an arbitrary number of descriptions is straight forward. We consider only the cases that leads to uniform (non-fractional) downsampling, i.e.\ reception of any single description, every other description (i.e.\ two out of four), or all four descriptions.

It can easily be seen that if we receive all four descriptions, the central distortion $d_c$ is given by the noise that falls within the in-band spectrum. In other words,
\begin{equation} \label{eq:dcK4}
d_c = \frac{1}{2\pi}\int_{-\pi/4}^{\pi/4} S_\epsilon(\omega) d\omega
\end{equation}
where $S_\epsilon(\omega)=|c(e^{j\omega})|^2\sigma_E^2$ denotes the power spectrum of the shaped noise. Similarly, when receveiving two out of four descriptions (i.e.\ one of the pair of descriptions (0,2) or (1,3)) the side distortion $d_2$ is given by
\begin{equation} \label{eq:d2K4}
d_2 = \frac{1}{2\pi}\int_{-\pi/4}^{\pi/4} S_\epsilon(\omega) d\omega + \frac{1}{2\pi}\int_{\frac{3}4\pi \leq |\omega| < \pi } S_\epsilon(\omega) d\omega
\end{equation}
where the latter term is due to aliasing (since we downsample by two without applying any anti-aliasing filter). Finally, if we receive only a single description and thereby downsample by four, the side distortion $d_1$ is given by the complete shaped noise spectrum, that is
\begin{equation} \label{eq:d1K4}
d_1 = \frac{1}{2\pi}\int_{-\pi}^{\pi} S_\epsilon(\omega) d\omega.
\end{equation}

\begin{figure}[th]
\psfrag{0}{\small $0$}
\psfrag{pi/4}{\small $\frac{\pi}{4}$}
\psfrag{pi/2}{\small $\frac{\pi}{2}$}
\psfrag{3pi/4}{\small $\frac{3\pi}{4}$}
\psfrag{pi}{\small $\pi$}
\psfrag{d0}{\small $\delta_0$}
\psfrag{d1}{\small $\delta_1$}
\psfrag{sd0d1}{\small $\delta_2=1/\sqrt{\delta_0\delta_1}$}
\psfrag{C(z)}{\small\hspace{-5mm} $|c(e^{j\omega})|^2\sigma_E^2$}
\psfrag{w}{\small $\omega$}
\begin{center}
\includegraphics[width=7cm]{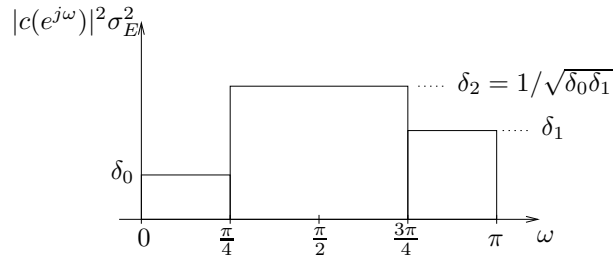}
\caption{An example of a shaped noise power spectrum $|c(e^{j\omega})|^2\sigma_E^2$ for $K=4$ descriptions.}
\label{fig:threeflatregions}
\end{center}
%\vspace{-3mm}
\end{figure}

Once again, we let $p\rightarrow \infty$ and take advantage of the frequency-domain interpretation, which we previously presented for the case of two descriptions. We divide the power spectrum of the shaped noise into three flat regions as shown in~Fig.~\ref{fig:threeflatregions}. The low frequency band (i.e.\ $|\omega|\leq \pi/4$) is of power $\delta_0$, the middle band (i.e.\ $\pi/4<|\omega|\leq 3\pi/4$) is of power $\delta_2$, and the high band (i.e.\ $3\pi/4< |\omega|< \pi$) is of power $\delta_1$.
With this choice of noise shaping, we guarantee that $c(z)$ is minimum phase simply by letting $\delta_2 = 1/\sqrt{\delta_0\delta_1}$ so that $\int_{-\pi}^{\pi}\log_2 S_\epsilon(\omega)d\omega = 0$.
From~(\ref{eq:dcK4}) --~(\ref{eq:d1K4}) it follows that\footnote{For clarity we have excluded the post multipliers, which are required for optimal reconstruction at general resolution. At high resolution conditions, the post multipliers become trivial and will not affect the distortions.}
\begin{equation} \label{eq:K4dc_pinf}
d_c = \frac{\sigma_E^2}{4}\delta_0,
\end{equation}
\begin{equation} \label{eq:K4d2_pinf}
\begin{split}
d_2 &= \frac{\sigma_E^2}{4}(\delta_0 + \delta_1)\\
 &= d_c + \frac{\sigma_E^2}{4}\delta_1
\end{split}
\end{equation}
and
\begin{equation} \label{eq:K4d1_pinf}
\begin{split}
d_1 &= \frac{\sigma_E^2}{4}(\delta_0 + \delta_1 + 2/\sqrt{\delta_0\delta_1})\\
& = d_2 + \frac{\sigma_E^2}{2\sqrt{\delta_0\delta_1}}.
\end{split}
\end{equation}

The description rate follows easily from previous results since the source is memoryless after downsampling. Specifically, it is easy to show that when using memoryless entropy coding the rate is given by
\begin{equation*}
\begin{split}
R &= \frac{1}{2}\log_2\left( \frac{ \sigma_X^2 + \sigma_E^2(\delta_0+\delta_1 + 2/\sqrt{\delta_0\delta_1})/4}{\sigma_E^2}\right) \\
& \approx \frac{1}{2}\log_2(\sigma_X^2/\sigma_E^2),
\end{split}
\end{equation*}
where the approximation becomes exact at high resolution.

It is worth emphasizing that in this example we have two controlling parameters, i.e.\ $\delta_0$ and $\delta_1$, where $\delta_0\leq 1$ and $\delta_0\delta_1\leq 1$. It is therefore possible to achieve almost arbitrary distortion ratios $d_1/d_2, d_1/d_c$ and $d_2/d_c$.
It was recently shown, see~\cite{zhang:2007}, that it is also possible to use several distortion controlling parameters in the source-splitting design of Chen et al.~\cite{chen:2006} and furthermore, by exploiting random binning, the achievable $K$-channel rate region of Pradhan et al.~\cite{pradhan:2004,puri:2005} can be achieved. Random binning can also be used to enlargen the rate region of the index-assignment based schemes, cf.~\cite{ostergaard:2006b,ostergaard:2007a}.

% is based on a single controlling parameter $\rho$ in the symmetric case. The parameter $\rho$ describes here the correlation between the noises of the $K$ descriptions just as was the case with Ozarow's solution for the two-channel problem.
% In order to increase the number of controlling parameters, it appears to be necessary to exploit the concept of binning as was done in the distributed MD approach of Pradhan et al.~\cite{pradhan:2004,puri:2005}.
%The index-assignment based schemes for $K\geq 2$ descriptions by \O stergaard et al.~\cite{ostergaard:2005c,ostergaard:2004b,ostergaard:2007a} also rely upon a single controlling parameter $N$ (in the symmetric case).
%Here $N$ describes the sublattice index of a nested sublattice. It is also possible to obtain additional controlling parameters for the index-assignment based schemes by using the binning approach of~\cite{pradhan:2004,puri:2005}, cf.~\cite{ostergaard:2006b,ostergaard:2007a}.

For the case of distributed source coding problems, e.g.\ the Wyner-Ziv problem, efficient binning schemes based on nested lattice codes have been proposed by Zamir et al.~\cite{zamir:2002}. However, these binning schemes are not (directly) applicable for the MD problem.\footnote{By making use of time-sharing, it is possible to apply the binning schemes presented in~\cite{zamir:2002} to the MD problem.} An alternative binning approach based on generalized coset codes has recently been proposed by Pradhan and Ramchandran~\cite{pradhan:2005}. It was indicated in~\cite{pradhan:2005} that the coset-based binning approach is applicable also for MD coding but the inherent rate loss was not addressed. Thus, the problem of designing efficient capacity achieving binning codes for the MD problem appears to be unsolved. From a practical point of view, it is therefore desirable to avoid binning. While the proposed MD design based on Delta-Sigma quantization avoids binning, we do not know whether there is a price to be paid in terms of rate loss.\footnote{Notice that we can reduce the coding rate by undersampling the signal so that the source spectrum will contain aliasing. 
As more descriptions are received, the lesser aliasing and a better reconstruction quality can be achieved. This stands in contrast to binning, where one can usually not reconstruct at all when too few descriptions are received.}
We Leave it as a topic of future research, to construct optimal reconstruction rules for the cases of non-uniform downsampling and furthermore addressing the issue whether the achievable $K$-channel rate-distortion region coincide with the one obtained by Pradhan et al.~\cite{pradhan:2004,puri:2005}.

%%%%%%%%%%%%%%%%%%%%%%%%%%%%%%%%%%%%%%%%%%%%%%%%%%%%%%%%%%%%%%%%%%%%%%%%%%%%%%%%%%%%%%%%%

\section{Conclusions and Discussion}
\label{sec:conclusion}
We proposed a symmetric two-channel MD coding scheme based on dithered Delta-Sigma quantization. We showed that for large vector quantizer dimension and large noise shaping filter order it was possible to achieve the symmetric two-channel MD rate-distortion function for a memoryless Gaussian source and MSE fidelity criterion. The construction was shown to be inherently symmetric in the description rate and there was therefore no need for source-splitting as were the case with existing related designs. It was shown that by simply increasing the oversampling ratio from two to $K$ it was possible to construct $K$ descriptions. 
%Moreover, the distortions resulting when reconstructing using distinct subsets of the $K$ descriptions could, in certain cases, be separately controlled via the noise shaping filter without the use of binning. 
The design of optimal reconstruction rules for $K>2$ descriptions was left as an open problem. 
Currently, we are working on extending the scheme to include prediction, in order to make it optimal for encoding sources with memory, without requiring entropy coders with memory, see~\cite{kochman:2008}.

% We would like to point out another interesting topic for future research which is that of extending
% the proposed design to sources with memory.
% One way to handle memory is to jointly entropy-code the quantizer outputs within each packet,
% as hinted in the end of the previous section.
% Yet, for Gaussian sources,
% it might be more advantageous to include a prediction loop to remove the source memory in line
% with what was done in~\cite{zamir:2007}.
% In this case, two feedback loops are utilized resulting in a nested predictive noise-shaping coder.

%%%%%%%%%%%%%%%%%%%%%%%%%%%%%%%%%%%%%%%%%%%%%%%%%%%%%%%%%%%%%%%%%%%%%%%%%%%%%%%%%%%%%%%%%%%%%%%%%%%%%%%%%%%%%%%%%%%

\appendix

\section*{Proof of Lemma~\ref{lem:dcpdc}}
%\begin{IEEEproof}%[Proof of Lemma~\ref{lem:dcpdc}]
We first find the central distortion (at high-resolution) through a 
time-domain approach and use the insights in order to find the optimal $\beta$, which will then lead to the central distortion at general resolution.

%
%\begin{lemma}\label{lem:mse}\label{lem:centraldist}
%Given the noise-shaping filter $(c_0,\cdots,c_p)$ of order $p\in \mathbb{Z}^+$, the central distortion, at high-resolution conditions, is given by
%\begin{equation}\label{eq:mse}
%d_c= \frac{\sigma_E^2}{2}\sum_{i=0}^{p}\sum_{j=0}^{p}
%\sinc\left(\frac{i-j}{2}\right)c_ic_j.
%\end{equation}
%\end{lemma}
%\begin{IEEEproof}
%
Let $\epsilon_n = \hat{x}_n - x_n$ be the error signal. Without loss of generality, we may view the upsampling operation followed by ideal lowpass filtering as an over-complete expansion of the source, where the infinite-dimensional analysis frame vectors with coefficients $\tilde{h}_{k,n}=\sinc(\frac{n-k}{2})$ are translated \sinc\ functions\footnote{The \sinc\ function is defined by
\begin{equation*}
\sinc(x) \triangleq \begin{cases}
\frac{\sin(\pi x)}{\pi x}, & x\neq 0 \\
1, & x=0.
\end{cases}
\end{equation*}
}. Thus, adopting the notation of~\cite{boufounos:2006}, we have that
\begin{equation*}
a_k = \sum_{n=-\infty}^{\infty} x_n\sinc\left(\frac{n-k}2\right)
\end{equation*}
and the synthesis filters are given by $h_{k,n}=\frac{1}{2}\sinc(\frac{n-k}{2})$, so that
\begin{equation*}
x_n = \frac{1}{2}\sum_{k=-\infty}^{\infty} a_k \sinc\left(\frac{n-k}2\right).
\end{equation*}

Since $\hat{a}_k = a_k + e_k +\sum_{i=1}^{p}c_ie_{k-i}$, the error $\epsilon_{n} = \hat{x}_{n} - x_n$ is given by
\begin{equation}\label{eq:error}
\epsilon_{n} = \sum_{k=-\infty}^{\infty}h_{k,n}\left(\sum_{i=0}^{p}c_ie_{k-i}\right).
\end{equation}
The (per sample) mean squared error (MSE) is (by use of~(\ref{eq:error})) given by
{\allowdisplaybreaks[4]
\begin{align} \notag
%\begin{split}
\mathbb{E}\epsilon^2_{n} &= \mathbb{E}\Bigg[\left(\sum_{k=-\infty}^{\infty}h_{k,n}\left(\sum_{i=0}^{p}c_iE_{k-i}\right)\right)^2\Bigg]\\ \notag
&=  \mathbb{E}\Bigg[\sum_{k=-\infty}^{\infty}\sum_{l=-\infty}^{\infty}h_{k,n}h_{l,n} \left(\sum_{i=0}^{p}c_iE_{k-i}\right) \\ \notag
&\quad\times\left(\sum_{i=0}^{p}c_iE_{l-i}\right)\Bigg] \\ \notag
&= \frac{1}{4} \mathbb{E}\Bigg[\sum_{k=-\infty}^{\infty}\sum_{l=-\infty}^{\infty}\sinc\left(\frac{n-k}{2}\right)\sinc\left(\frac{n-l}{2}\right)\\ \notag
&\quad\times \left(\sum_{i=0}^{p}c_iE_{k-i}\right)\left(\sum_{i=0}^{p}c_iE_{l-i}\right)\Bigg] \\ \notag
&= \frac{1}{4}\sum_{k=-\infty}^{\infty}\sum_{l=-\infty}^{\infty}\sinc\left(\frac{n-k}{2}\right)\sinc\left(\frac{n-l}{2}\right) \sum_{i=0}^{p}\sum_{j=0}^{p} c_ic_j \\ \notag
&\quad\times \mathbb{E}[E_{k-i}E_{l-j}] \\ \notag
&= \frac{1}{4}\sum_{k=-\infty}^{\infty}\sinc\left(\frac{n-k}{2}\right) \sum_{i=0}^{p}\sum_{j=0}^{p} c_ic_j \\ \notag &\quad\times\mathbb{E}\Bigg[E_{k-i}\sum_{l=-\infty}^{\infty}\sinc\left(\frac{n-l}{2}\right)E_{l-j}\Bigg] \\ \notag
&\overset{(a)}{=} \frac{1}{4}\sum_{k=-\infty}^{\infty}\sinc\left(\frac{n-k}{2}\right) \sum_{i=0}^{p}\sum_{j=0}^{p} c_ic_j \\ \notag
&\quad\times\mathbb{E}\Bigg[E_{k-i}^2\sinc\left(\frac{n-k-i+j}{2}\right)\Bigg] \\ \label{eq:mse_hr1}
&\overset{(b)}{=} \frac{\sigma_E^2}{2}\sum_{i=0}^{p}\sum_{j=0}^{p}
\sinc\left(\frac{i-j}{2}\right)c_ic_j,
%\end{split}
\end{align}
}
where $(a)$ follows from the fact that $\mathbb{E}E_{k-i}E_{l-j}$ is non-zero only when $l-j=k-i$ which implies that $l=k-i+j$ and $(b)$ is due to the following property of the \sinc\ function~\cite{mcnamee:1971}
\begin{equation*}
\sum_{k=-\infty}^{\infty}\sinc\left(c_0 - \frac{k}{r}\right)\sinc\left(c_0 - \frac{k-c_1}{r}\right) = r\,\sinc\left(\frac{c_1}{r}\right).
\end{equation*}

Let $U$ denote the reconstruction before the central post multiplier $\beta$, i.e.\ $U$ is the variable obtained by first lowpass filtering $A_k$ and then 
downsampling by two. It follows immediately that $\mathbb{E}U^2 = \sigma_X^2 + \frac{1}{2\pi}\int_{|\omega|<\frac{\pi}2}|c(e^{j\omega})|^2\sigma_E^2d\omega = \sigma_X^2 + \sigma_E^2P_{d_c}$. Furthermore, from~(\ref{eq:mse_hr1}) it can be seen that 
\begin{equation}\label{eq:EU0U1}
\mathbb{E}U^2 = \sigma_X^2 + \frac{\sigma_E^2}{2}\sum_{i=0}^{p}\sum_{j=0}^{p}\sinc\left(\frac{i-j}2\right)c_ic_j.
\end{equation}
so using that $\mathbb{E}[X|U]=\beta U$ yields
\begin{align}\notag 
\beta &= \frac{\sigma_X^2}{\sigma_X^2 + \frac{\sigma_E^2}{2}\sum_{i=0}^p\sum_{j=0}^{p}\sinc(\frac{i-j}{2})c_ic_j} \\  \label{eq:opt_beta}
&=\frac{\sigma_X^2}{\sigma_X^2 + \sigma_E^2P_{d_c}}.
\end{align}
The central distortion at general resolution now follows by inserting $\beta$ (\ref{eq:opt_beta}) into~(\ref{eq:beforecentralpost}), which leads to~(\ref{eq:dcPdc}).

We will now derive the side distortion. First notice that since we only receive either all odd samples or all even samples, we should only sum over terms in~(\ref{eq:mse_hr1}) where the lag $|i-j|$ is even. 
However, all cross-terms, $c_ic_j, i\neq j$, vanish since $\sinc(x/2)=0$ for $x=\pm 2,\pm 4,\dotsc$, so only the $p+1$ auto-terms, $c_i^2, i=0,\dotsc, p$, contribute to the distortion. In addition, we make use of the following property of the \sinc\ function~\cite{mcnamee:1971}
\begin{equation}\label{eq:sincprop1}
\sum_{k=-\infty}^{\infty}\sinc\left(\frac{xk}{r}\right)\sinc\left(\frac{xk-c}{r}\right) = \frac{r}{x}\sinc\left(\frac{c}{r}\right).
\end{equation}
With this, it follows that the high-resolution side distortion $d_s^{hr}$ is given by
\begin{equation*}
d_s^{hr} = \sigma_E^2\sum_{i=0}^{p}c_i^2 = \sigma_E^2P_{d_s}.
\end{equation*}
At this point, we let $U_i$ denote the reconstruction before the multiplier $\alpha$ such that $\hat{X}_i = \alpha U_i, i=0,1$. It should be clear that $\mathbb{E}XU_i=\sigma_X^2$.\footnote{The even samples are noisy versions of $X$ where the noise is independent of $X$. The odd samples are noisy and phase shifted versions of $X$. However, the phase shift is corrected by the all-pass filter $h_p(z)$ before the post multiplier. Thus, $\mathbb{E}XU_i=\sigma_X^2, i=0,1$.}  
Recall that the autocorrelation of the even lags of $U_i$ vanish so that 
\begin{equation}\label{eq:Ui2}
\mathbb{E}U_i^2 = \sigma_X^2 + \sigma_E^2\sum_{i=0}^{p}c_i^2 = \sigma_X^2 + \sigma_E^2P_{d_s}
\end{equation}
Since, $\mathbb{E}[X|U_i] = \alpha U_i$, it follows that 
\begin{equation}\label{eq:opt_alpha}
\alpha =\frac{\sigma_X^2}{\sigma_X^2 + \sigma_E^2\sum_{i=0}^{p}c_i^2}.
\end{equation}
Inserting~(\ref{eq:opt_alpha}) into~(\ref{eq:beforesidepost}) leads to~(\ref{eq:dsPds}), which is the side distortion at general resolution.
This completes the proof. \hfill \IEEEQED
%\end{IEEEproof}

\section*{Proof of Lemma~\ref{lem:opt_uniq}}
%\begin{IEEEproof}[Proof of Lemma~\ref{lem:opt_uniq}]
A minimum-phase filter $H(z)$ with power spectrum $S(f)= |H(e^{j2\pi f})|^2$, $-1/2<f\leq 1/2$ satisfies
\begin{equation*}
e^{\int_{-1/2}^{1/2}\ln S(f) df} = |h_0|^2
\end{equation*}
where $h_0$ is the zero-tap of the filter. It is also known that the zero-tap of a minimum-phase filter is strictly larger than the zero-tap of a non-minimum-phase filter having the same power spectrum~\cite{oppenheim:1989}. Thus, for an arbitrary filter $H(z)$ with power spectrum $S(f)$ and zero-tap $h_0$
\begin{equation*}
e^{\int_{-1/2}^{1/2}\ln S(f) df} \geq |h_0|^2
\end{equation*}
with equality if and only if $H(z)$ is minimum phase.
Furthermore, from the geometric-arithmetic means inequality it can be shown that
\begin{align} \label{eq:spectralproduct}
2\sqrt{ \int_{|f|\leq 1/4}\!\!\! \!\!\!S(f) df \int_{1/4 < |f| < 1/2}\!\!\!\!\!\! S(f) df } 
&\geq  e^{\int_{-1/2}^{1/2}\ln S(f) df}\\ \label{eq:spectralproduct1}
&\geq 1
\end{align}
where we used the fact that in our case the filter is monic, so $h_0=1$ 
and where we have equality in~(\ref{eq:spectralproduct}) and~(\ref{eq:spectralproduct1}) 
if and only if the filter $H(z)$ is minimum phase and the power spectrum consists of two flat regions; $S(f)=\delta^{-1}$ for $|f|\leq 1/4$ and $S(f)=\delta$ for $1/4<|f|<1/2$. 
Let us now fix the energy ratio $P_{d_s}/P_{d_c}=\gamma$, where $1\leq \gamma\in \mathbb{R}$,
$P_{d_c} = \int_{|f|\leq 1/4} S(f) df$ and $P_{d_s} = \int_{|f|\leq 1/4} S(f) df + \int_{1/4 < |f| < 1/2} S(f) df$. With this, it follows that
\begin{equation}\label{eq:HPoverLP}
\frac{\int_{1/4 < |f| < 1/2} S(f) df}{\int_{|f|\leq 1/4} S(f) df } = \gamma -1.
\end{equation}
Using~(\ref{eq:HPoverLP}) in the left-hand-side of~(\ref{eq:spectralproduct}) leads to the following two inequalities
\begin{equation}\label{eq:lb_dc}
P_{d_c} \geq \frac{1}{2}\frac{1}{\sqrt{\gamma-1}} = \frac{1}{2}\delta^{-1}
\end{equation}
and
\begin{equation}\label{eq:lb_ds}
P_{d_s} \geq \frac{1}{2}\sqrt{\gamma-1} + \frac{1}{2}\frac{1}{\sqrt{\gamma-1}} = \frac{1}{2}(\delta+\delta^{-1})
\end{equation}
where we have equality in both~(\ref{eq:lb_dc}) and~(\ref{eq:lb_ds}) (at the same time) if and only if the filter is minimum phase and the power spectrum is a two-step function, i.e.\ it has constant power $\delta^{-1}=1/\sqrt{\gamma-1}$ through-out the lowpass band and $\delta$ through-out the highpass band.
%\footnote{In fact we only require that $\einf S(f) = \delta^{-1}$ for $|f|\leq \frac{1}{4}$ and $\esup S(f)=\delta$ for $\frac{1}4< |f|<\frac{1}2$.} 

At this point we let $\alpha= \frac{\sigma_X^2}{\sigma_X^2 + \sigma_E^2P_{d_s}}$ and $\beta=\frac{\sigma_X^2}{\sigma_X^2 + \sigma_E^2P_{d_c}}$ from which it can be shown that the distortions at general resolution are given by
\begin{equation}\label{eq:dcPdc1}
d_c = \frac{\sigma_X^2 \sigma_E^2P_{d_c}}{\sigma_X^2 + \sigma_E^2P_{d_c}}
\end{equation}
and
\begin{equation}\label{eq:dsPds1}
d_s = \frac{\sigma_X^2 \sigma_E^2P_{d_s}}{\sigma_X^2 + \sigma_E^2P_{d_s}}.
\end{equation}
Inserting the lower bounds of~(\ref{eq:lb_dc}) and~(\ref{eq:lb_ds})
into~(\ref{eq:dcPdc1}) and~(\ref{eq:dsPds1}) yields Ozarow's symmetric rate-distortion function (see~(\ref{eq:centralpost}) and (\ref{eq:sidedistpost})). Moreover, (\ref{eq:dcPdc1}) and~(\ref{eq:dsPds1}) are strictly increasing in $P_{d_c}$ and $P_{d_s}$, respectively. 
Thus, for a fixed ratio $\gamma$, any other spectrum than the two-step $S(f)$ given above must necessarily lead to a greater distortion.
To complete the proof, we remark that in order to have such an ideal brick-wall power spectrum, the order of the filter must necessarily be infinite.  \hfill \IEEEQED
%\end{IEEEproof}

\section*{Acknowledgment}
The authors would like to thank the referees and the associate editor for their valuable comments and suggestions which aided in clarifying the technical presentation. 
The authors would also like to thank Y.\ Kochman for providing useful comments on a draft version of this paper. 

%\IEEEtriggeratref{42}
\bibliographystyle{ieeetr}

\begin{thebibliography}{10}
\bibitem{candy:1992}
J.~C. Candy and G.~C. Temes, eds., {\em Oversampling Delta-Sigma Data
  Converters: Theory, Design and Simulation}.
\newblock {IEEE} Press, 1992.

\bibitem{elgamal:1982}
A.~A.~E. Gamal and T.~M. Cover, ``Achievable rates for multiple descriptions,''
  {\em IEEE Trans. Inf. Theory}, vol.~IT-28, pp.~851 -- 857, November 1982.

\bibitem{ozarow:1980}
L.~Ozarow, ``On a source-coding problem with two channels and three
  receivers,'' {\em Bell System Technical Journal}, vol.~59, pp.~1909 -- 1921,
  December 1980.

\bibitem{zamir:1999}
R.~Zamir, ``Gaussian codes and shannon bounds for multiple descriptions,'' {\em
  IEEE Trans. Inf. Theory}, vol.~45, pp.~2629 -- 2636, November 1999.

\bibitem{zamir:2000}
R.~Zamir, ``Shannon type bounds for multiple descriptions of a stationary
  source,'' {\em Journal of Combinatorics, Information and System Sciences},
  pp.~1 -- 15, December 2000.

\bibitem{venkataramani:2003}
R.~Venkataramani, G.~Kramer, and V.~K. Goyal, ``Multiple description coding
  with many channels,'' {\em IEEE Trans. Inf. Theory}, vol.~49, pp.~2106 --
  2114, September 2003.

\bibitem{pradhan:2004}
S.~S. Pradhan, R.~Puri, and K.~Ramchandran, ``{$n$}-channel symmetric multiple
  descriptions--part {I}: {$(n,k)$} source-channel erasure codes,'' {\em IEEE
  Trans. Inf. Theory}, vol.~50, pp.~47 -- 61, January 2004.

\bibitem{puri:2005}
R.~Puri, S.~S. Pradhan, and K.~Ramchandran, ``{$n$}-channel symmetric multiple
  descriptions- part {II}: An achievable rate-distortion region,'' {\em IEEE
  Trans. Inf. Theory}, vol.~51, pp.~1377 -- 1392, April 2005.

\bibitem{vaishampayan:1993}
V.~A. Vaishampayan, ``Design of multiple description scalar quantizers,'' {\em
  IEEE Trans. Inf. Theory}, vol.~39, pp.~821 -- 834, May 1993.

\bibitem{vaishampayan:2001}
V.~A. Vaishampayan, N.~J.~A. Sloane, and S.~D. Servetto, ``Multiple-description
  vector quantization with lattice codebooks: Design and analysis,'' {\em IEEE
  Trans. Inf. Theory}, vol.~47, pp.~1718 -- 1734, July 2001.

\bibitem{diggavi:2002}
S.~N. Diggavi, N.~J.~A. Sloane, and V.~A. Vaishampayan, ``Asymmetric multiple
  description lattice vector quantizers,'' {\em IEEE Trans. Inf. Theory},
  vol.~48, pp.~174 -- 191, January 2002.

\bibitem{ostergaard:2004b}
J.~{\O}stergaard, J.~Jensen, and R.~Heusdens, ``{$n$}-channel
  entropy-constrained multiple-description lattice vector quantization,'' {\em
  IEEE Trans. Inf. Theory}, vol.~52, pp.~1956 -- 1973, May 2006.

\bibitem{ostergaard:2007a}
J.~{\O}stergaard, {\em {M}ultiple-description lattice vector quantization}.
\newblock PhD thesis, Delft University of Technology, Delft, The Netherlands,
  June 2007.
\newblock Available online: http://arxiv.org/abs/0707.2482.

\bibitem{dayan:2002}
Y.~Frank-Dayan and R.~Zamir, ``Dithered lattice-based quantizers for multiple
  descriptions,'' {\em IEEE Trans. Inf. Theory}, vol.~48, pp.~192 -- 204,
  January 2002.

\bibitem{chen:2006}
J.~Chen, C.~Tian, T.~Berger, and S.~S. Hemami, ``Multiple description
  quantization via {G}ram-{S}chmidt orthogonalization,'' {\em IEEE Trans. Inf.
  Theory}, vol.~52, pp.~5197 -- 5217, December 2006.

\bibitem{gersho:1979}
A.~Gersho, ``Asymptotically optimal block quantization,'' {\em IEEE Trans. Inf.
  Theory}, vol.~{IT}-25, pp.~373 -- 380, July 1979.

\bibitem{jayant:1981}
N.~S. Jayant, ``Subsampling of a {DPCM} speech channel to provide two
  self-contained, half-rate channels,'' {\em Bell Syst. Tech. Jour.}, vol.~60,
  pp.~501 -- 509, April 1981.

\bibitem{costa:2005}
O.~L.~V. Costa, M.~D. Fragoso, and R.~P. Marques, {\em Discrete-Time Markov
  Jump Linear Systems}.
\newblock Springer, February 2005.

\bibitem{chou:1999}
P.~A. Chou, S.~Mehrotra, and A.~Wang, ``Multiple description decoding of
  overcomplete expansions using projections onto convex sets,'' in {\em Proc.
  Data Compression Conf.}, pp.~72 -- 81, March 1999.

\bibitem{goyal:2001b}
V.~Goyal, J.~Kova\v{c}evi\'{c}, and J.~Kelner, ``Quantized frame expansions
  with erasures,'' {\em Journal of Appl. and Comput. Harmonic Analysis},
  vol.~10, pp.~203--233, May 2001.

\bibitem{dragotti:2001}
P.~L. Dragotti, J.~Kova\v{c}evi\'{c}, and V.~K. Goyal, ``Quantized oversampled
  filter banks with erasures,'' in {\em Proc. Data Compression Conf.}, pp.~173
  -- 182, March 2001.

\bibitem{kovacevic:2002}
J.~Kova\v{c}evi\'{c}, P.~L. Dragotti, and V.~K. Goyal, ``Filter bank frame
  expansions with erasures,'' {\em IEEE Trans. Inf. Theory}, vol.~48, pp.~1439
  -- 1450, June 2002.

\bibitem{bolcskei:2001}
H.~B\"{o}lcskei and F.~Hlawatsch, ``Noise reduction in oversampled filter banks
  using predictive quantization,'' {\em IEEE Trans. Inf. Theory}, vol.~47,
  pp.~155 -- 172, January 2001.

\bibitem{boufounos:2005}
P.~T. Boufounos and A.~V. Oppenheim, ``Compensation of coefficient erasures in
  frame representations,'' in {\em Proc. IEEE Int. Conf. Acoustics, Speech, and
  Signal Processing}, vol.~3, pp.~848 -- 851, May 2006.

\bibitem{boufounos:2006}
P.~T. Boufounos and A.~V. Oppenheim, ``Quantization noise shaping on arbitrary
  frame expansions,'' {\em {EURASIP} Journal on Applied Signal Processing},
  vol.~2006, pp.~1 -- 12, 2006.

\bibitem{zamir:1992}
R.~Zamir and M.~Feder, ``On universal quantization by randomized
  uniform/lattice quantizer,'' {\em IEEE Trans. Inf. Theory}, vol.~38, pp.~428
  -- 436, March 1992.

\bibitem{conway:1999}
J.~H. Conway and N.~J.~A. Sloane, {\em Sphere packings, Lattices and Groups}.
\newblock Springer, 3rd~ed., 1999.

\bibitem{zamir:1996}
R.~Zamir and M.~Feder, ``On lattice quantization noise,'' {\em IEEE Trans. Inf.
  Theory}, vol.~42, pp.~1152 -- 1159, July 1996.

\bibitem{massey:1990}
J.~L. Massey, ``Causality, feedback and directed information,'' in {\em Proc.
  Int. Symp. on Info. Th. \& its Appls.}, pp.~303 -- 305, November 1990.

\bibitem{zamir:2007}
R.~Zamir, Y.~Kochman, and U.~Erez, ``Achieving the {G}aussian rate-distortion
  function by prediction,'' {\em IEEE Trans. Inf. Theory}, vol.~54, pp.~3354 --
  3364, July 2008.

\bibitem{zamir:1996b}
R.~Zamir and M.~Feder, ``Information rates of pre/post-filtered dithered
  quantizers,'' {\em IEEE Trans. Inf. Theory}, vol.~42, pp.~1340 -- 1353,
  September 1996.

\bibitem{tewksbury:1978}
S.~K. Tewksbury and R.~W. Hallock, ``Oversampled, linear predictive and
  noise-shaping coders of order ${N>1}$,'' in {\em Oversampling Delta-Sigma
  Data Converters: Theory, Design and Simulation} (J.~C. Candy and G.~C. Temes,
  eds.), pp.~139 -- 149, {IEEE} Press, 1992.

\bibitem{vaishampayan:1998c}
V.~A. Vaishampayan, J.-C. Batllo, and A.~Calderbank, ``On reducing granular
  distortion in multiple description quantization,'' in {\em Proc. IEEE Int.
  Symp. Information Theory}, p.~98, August 1998.

\bibitem{markel:1976}
J.~D. Markel and A.~H. Gray, {\em Linear prediction of speech}.
\newblock New {Y}ork: {S}pringer-{V}erlag, 1976.

\bibitem{bachman:2000}
G.~Bachman, L.~Narici, and E.~Beckenstein, {\em Fourier and wavelet analysis}.
\newblock Springer, 2000.

\bibitem{vetterli:2001}
M.~Vetterli, ``Wavelets, approximation, and compression,'' {\em IEEE Signal
  Processing Magazine}, vol.~18, pp.~59 -- 73, September 2001.

\bibitem{makhoul:1975}
J.~Makhoul, ``Linear prediction: A tutorial review,'' {\em Proceedings of the
  IEEE}, vol.~63, pp.~561 -- 580, April 1975.

\bibitem{kochman:2008}
Y.~Kochman, J.~{\O}stergaard, and R.~Zamir, ``Noise-shaped predictive coding
  for multiple descriptions of a colored {G}aussian source,'' in {\em Proc.
  Data Compression Conf.}, pp.~362 -- 371, March 2008.

\bibitem{zhang:2007}
X.~Zhang, J.~Chen, S.~B. Wicker, and T.~Berger, ``Successive coding in
  multiuser information theory,'' {\em IEEE Trans. Inf. Theory}, vol.~53,
  pp.~2246 -- 2254, June 2007.

\bibitem{ostergaard:2006b}
J.~{\O}stergaard, R.~Heusdens, and J.~Jensen, ``Source-channel erasure codes
  with lattice codebooks for multiple description coding,'' in {\em Proc. IEEE
  Int. Symp. Information Theory}, pp.~2324 -- 2328, July 2006.

\bibitem{zamir:2002}
R.~Zamir, S.~Shamai, and U.~Erez, ``Nested linear/lattice codes for structured
  multiterminal binning,'' {\em IEEE Trans. Inf. Theory}, vol.~48, pp.~1250 --
  1276, June 2002.

\bibitem{pradhan:2005}
S.~S. Pradhan and K.~Ramchandran, ``Generalized coset codes for distributed
  binning,'' {\em IEEE Trans. Inf. Theory}, vol.~51, pp.~3457 -- 3474, October
  2005.

\bibitem{mcnamee:1971}
J.~McNamee, F.~Stenger, and E.~L. Whitney, ``Whittaker's cardinal spline
  function in retrospect,'' {\em American Mathematical Society}, vol.~25,
  pp.~141 -- 154, January 1971.

\bibitem{oppenheim:1989}
A.~V. Oppenheim and R.~W. Schafer, {\em Discrete-time signal processing}.
\newblock Prentice {H}all, 1989.

\end{thebibliography}

\begin{biographynophoto}{Jan {\O}stergaard}
(S'98-M'99)
received the M.Sc.\ degree in electrical engineering from Aalborg University, Aalborg, Denmark, in 1999 and the Ph.D.\ degree (with cum laude) in electrical engineering from Delft University of Technology, Delft, The Netherlands, in 2007. 
From 1999 to 2002, he worked as an R\&D engineer at ETI A/S, Aalborg, Denmark, and from 2002 to 2003, he worked as an R\&D engineer at ETI Inc., Virginia, United States. 
Between September 2007 and June 2008, he worked as a post-doctoral researcher in the Centre for Complex Dynamic Systems and Control, School of Electrical Engineering and Computer Science, The University of Newcastle, NSW, Australia. 
He has also been a visiting researcher at Tel Aviv University, Tel Aviv, Israel. 
He is currently a post-doctoral researcher at Aalborg University, Aalborg, Denmark. 
Dr.\ {\O}stergaard has received a Danish Independent Research Council's Young Researcher's Award and a fellowship from the Danish Research Council for Technology and Production Sciences.
\end{biographynophoto}

% insert where needed to balance the two columns on the last page
%\newpage

\begin{biographynophoto}{Ram Zamir}
(S'89-M'95-SM'99) was born in Ramat-Gan, Israel, in 1961. 
 He received the B.Sc., M.Sc.~(summa cum laude), and D.Sc. (with distinction) degrees
 from Tel-Aviv University, Tel-Aviv, Israel, in 1983, 1991, and 1994, respectively,
 all in electrical engineering.
 In the years 1994-1996, he spent a postdoctoral period at Cornell University,
 Ithaca, NY, and at the University of California, Santa Barbara. During 2002,
 he spent a sabbatical year at the Massachusetts Institute of Technology (MIT),
 Cambridge. Since 1996, he has been with the Department of Electrical Engineering-
 Systems at Tel-Aviv University. He has been consulting in the areas of
 radar and communications, mainly in developing algorithms and in the design
 of signals, and has been teaching information theory, data compression, random
 processes, communications systems, and communications circuits at Tel-Aviv
 University. His research interests include information theory, communication
 and remote sensing systems, and signal processing.
 Dr.\ Zamir received the Israel Ministry of Communications Award in 1993,
 and the Wolfson Post-Doctoral Research Award in 1994, and visited the Technical
 University of Budapest, Budapest, Hungary, under a joint program of the
 Israeli and Hungarian academies of science in summer 1995. He served as an Associate
 Editor for Source Coding for the IEEE TRANSACTIONS ON INFORMATION
 THEORY (2001-2003), and headed the Information Theory Chapter of the Israeli
 IEEE Society (2000-2005).
\end{biographynophoto}

% You can push biographies down or up by placing
% a \vfill before or after them. The appropriate
% use of \vfill depends on what kind of text is
% on the last page and whether or not the columns
% are being equalized.

\vfill

% Can be used to pull up biographies so that the bottom of the last one
% is flush with the other column.
%\enlargethispage{-5in}

% that's all folks
\end{document}